\documentclass[12pt,onecolumn]{IEEEtran}
\usepackage{color}
\usepackage{multirow}
\usepackage{subfig}
\ifCLASSINFOpdf   
   \usepackage[pdftex]{graphicx}     
   \graphicspath{{./}{img/}{../pdf/}{../jpeg/}{./image/}{./img/}}    
   \DeclareGraphicsExtensions{.pdf,.jpeg,.png,.jpg}   
 \else   
   \usepackage[dvips]{graphicx}   
   \graphicspath{{../eps/}}    
\fi

\usepackage{wrapfig}
\usepackage{lscape}
\usepackage{rotating}
\usepackage{capt-of}
\usepackage{fancyhdr}

\fancypagestyle{empty}{%
  \fancyhf{}
  \fancyhead[L]{\small {\color{blue} This is the accepted version of an article accepted for publication in \textit{IEEE Transactions on Fuzzy Systems\\Copyright @ 2016 IEEE}}}
}

\definecolor{Orange}{rgb}{1,0.5,0}

\newtheorem{xmpl}{Example} 

\begin{document}

\title{Using Fuzzy Logic to Leverage HTML Markup for Web Page Representation}
\author{
  Alberto P. Garc\'{i}a-Plaza$^1$, V\'{i}ctor Fresno$^1$, Raquel Mart\'{i}nez$^1$ and Arkaitz Zubiaga$^2$\\
  $^1$ NLP\&IR Group at UNED, Spain\\
  $^2$ University of Warwick, UK\\
  \{alpgarcia,vfresno,raquel\}@lsi.uned.es, arkaitz@zubiaga.org
}

\markboth{IEEE Transactions on Fuzzy Systems}%
{Shell \MakeLowercase{\textit{et al.}}: Bare Advanced Demo of IEEEtran.cls for Journals}

\IEEEcompsoctitleabstractindextext{%
\begin{abstract}
 The selection of a suitable document representation approach plays a crucial role in the performance of a document clustering task. Being able to pick out representative words within a document can lead to substantial improvements in document clustering. In the case of web documents, the HTML markup that defines the layout of the content provides additional structural information that can be further exploited to identify representative words. In this paper we introduce a fuzzy term weighing approach that makes the most of the HTML structure for document clustering. We set forth and build on the hypothesis that a good representation can take advantage of how humans skim through documents to extract the most representative words. The authors of web pages make use of HTML tags to convey the most important message of a web page through page elements that attract the readers' attention, such as page titles or emphasized elements. We define a set of criteria to exploit the information provided by these page elements, and introduce a fuzzy combination of these criteria that we evaluate within the context of a web page clustering task. Our proposed approach, called Abstract Fuzzy Combination of Criteria (AFCC), can adapt to datasets whose features are distributed differently, achieving good results compared to other similar fuzzy logic based approaches and TF-IDF across different datasets.
\end{abstract}

}

\maketitle

\IEEEdisplaynotcompsoctitleabstractindextext
\IEEEpeerreviewmaketitle

\thispagestyle{empty}

\section{Introduction}

\IEEEPARstart{A}{ccess} to and retrieval of web documents in large collections can be substantially eased when the documents are properly clustered into topics. The organization of documents into clusters then facilitates focusing search on the topic or topics of interest, shrinking down the large collection to smaller sets of topically related resources. While a body of research has studied clustering of web documents, little attention has been paid to the improvement of document representation techniques and the definition of robust term weighting functions.

We are interested in the study of document representation techniques based on fuzzy logic that can generalize across datasets when the purpose is to group documents by topic in the absence of category information.
This can be particularly useful in cases where new categories can emerge, so that the system should be able to accommodate its clustering process to be able to find these new categories with the information extracted from the documents. 

Prior to the clustering process, document representation plays a very important role in web page clustering, and constitutes the central point of research of this work. 
In the document representation phase we choose the characteristics of the document that we consider useful, and assess how this information could be exploited. 

The textual content is often used for the representation of web pages, given that it is readily available and is easy to process; however, an unweighted bag-of-words representation of the content does not always lead to optimal results. Interestingly, the content of an HTML document is structured in tags, providing additional clues on how different parts of the content differ from one another, and ultimately affecting its visual presentation \cite{qi_webpage_2009}. The HTML structure of a web document can be further exploited to identify the most representative words within its content.
We pay special attention to document contents, introducing a representation that makes the most of information inherent to the document. 
Hence, we set out to delve into the study of approaches that garner the additional information that HTML tags provide for improved representation of web documents. Moreover, we also look into the use of additional context information, using anchor texts pointing to web pages, as well as statistics inferred from the whole collection. We assess the suitability of using these additional characteristics for web document representation in a clustering task.

We make use of a fuzzy system, as a flexible solution that enables to handle the importance of the different characteristics of web pages. For instance, the titles of web pages can often be deemed rhetorical, where some words are very representative of its content, but other words are solely used to embellish the language. When considering frequency in titles within a linear combination of criteria in order to identify the most important words within a document, these words would get a high importance value, which would not correspond with their real importance to describe the content of the page, since they are only embellishing the language. In these linear combinations, when a word is important with respect to a single criterion, the corresponding component will have a value which will always be added to the importance of the word in the document, regardless of the importance corresponding of the rest of the components. On the contrary, by using fuzzy logic it is possible to define related conditions, e.g., a word should appear in the title and emphasized or within specific parts of the document to be considered important. In the same way, if a word appears in the title but not in other criteria, then we could consider that word less important. 
Here we delve into the use of fuzzy logic for the purposes of exploiting these characteristics of web pages.

Building on the state-of-the-art unsupervised fuzzy logic approach for HTML document representation \cite{Fresno06}, known as Fuzzy Combinations of Criteria (FCC), we propose three alternative approaches, namely EFCC, AddFCC, and AFCC. We perform the evaluation of these and additional baseline approaches over three benchmark web page collections through a clustering task using the well-known Cluto library \cite{karypis03}. Our proposed approach AFCC, which more suitably adapts to datasets with different characteristics, consistently outperforms the other approaches on the three datasets under study. AFCC provides a flexible, straightforwardly applicable approach that makes the most of the structure and content of HTML documents for web mining purposes.

In what follows, we provide background on the task of web page representation, followed by a summary of previous work in the literature as well as their relevance to our work in Section \ref{sec:related-work}. We move on then to the experimentation, describing first the experimental settings in Section \ref{sec:experimental-framework}, introducing and evaluating two new approaches, AddFCC and EFCC, to improve the existing FCC approach in Section \ref{sec:improving-combination}, and further studying the analysis and tuning of membership functions through the so-called AFCC in Section \ref{sec:afcc}. We outline the contributions of our work and conclude the paper in Section \ref{sec:discussion}.

\section{Background}
\label{sec:background}

The document representation process can be split into three stages: (1) selection of feature sources, (2) weighing of those features, and (3) dimensionality reduction. Throughout this paper we delve into these three stages, paying close attention at how to model a term weighing function.

Within the \textbf{selection of feature sources}, the information that needs to be represented within each document is picked, e.g., plain textual content, titles, or hyperlinks. There are mainly three different approaches. First, \textit{content based}, which make use of the textual content of documents. This kind of approaches were initially developed for document retrieval in static collections, but with the popularity of the Internet, they have also been adapted to the Web. Further exploiting the characteristics of the Web, the textual content of the documents has also been enhanced with the information provided by HTML tags about document formatting, page structure, visual aspects, etc. Second, \textit{link based}, which take advantage of the link structure among the pages in the collections. It considers hyperlinks as citations between pages. When two documents have many incoming links in common, or both documents have outgoing links to a similar set of documents, then the documents are likely related. Third and last, the \textit{hybrid approach}, which combines features from the textual content of the document and from the context of the page. Here context can include not only hyperlinks or anchor texts, but also other information sources, such as information inferred from the entire collection, or definitions extracted from external resources such as Wikipedia.

In the subsequent step of \textbf{weighing features}, each feature is assigned a weight in each document, the weight being representative of the feature's importance in the document. There are different elements that can determine the importance of a word within the document. One can then define a set of criteria to make the most of the different elements when it comes to improving the document representation. The initial hypothesis of the present work lies in that a good representation should take advantage of how humans skim through web documents to pick out salient words. For example, some words are explicitly highlighted with specific HTML tags. Then, if one wants to determine the importance of a word in a document, in addition to the rather straightforward frequency of the word in the document, one can also take advantage of these highlighted words as a signal that conveys the remarkable importance of the word.

In the final \textbf{dimensionality reduction step}, useless features are removed by keeping the document's most representative features, which makes it more efficient to be handled computationally.

\section{Related Work}
\label{sec:related-work}

There have been multiple attempts at exploiting the structure of web pages to maximize understanding of their contents for different purposes. Kwon and Lee \cite{kwon03text} aimed to classify web sites by using not only their home pages, but also the content of pages that linked to the home page of each site. Their weighing scheme to establish term importance takes into account different HTML tags such as titles, headlines, and boldfaced texts, to identify the most representative words in a web page. They show that the use of the extended set of pages boosts the performance with respect to the ordinary classifier using only the home pages. Golub and Ard \cite{golub05importance} studied how setting the importance of different parts of a web page could have an impact on the outcome of a web page classification task. They classified a set of $1,003$ web pages based on titles, headings, metadata, and text. As a single feature, they found the titles to be the most useful; however, since not all web pages have titles, they found that combining all features leads to the best overall performance. In an earlier work, making use of the link structure among documents, Fisher and Everson \cite{Fisher2003} analyzed the usefulness of links for web page classification tasks. They conclude that links may be useful, but it depends on link density and quality.

Besides links and anchor texts, other kinds of information have also been exploited over the years. For instance, 
Kovacevic et al. \cite{KOVACEVIC2004}, Shih and Karger \cite{Shih2004}, Bohunsky and Gatterbauer \cite{bohunsky2010} and Bartik \cite{bartik10}, or more recently Herzog et al. \cite{herzog2013}, have used the visual appearance of a web page, after rendering its content in a browser, for the purposes of representing web documents. Another work along these lines is that performed by Gasparetti et al. \cite{gasparettiMS14}, which describes an approach based on the implicit signal that can be captured through web browsing interactions, defining a DOM-based representation of visited pages. While these approaches might be handy for systems that exploit the visual appearance of web pages, our objective instead is to avoid reliance on the visual rendering by solely exploiting the HTML structure. 

Information from external knowledge bases such as Wikipedia has also been exploited by others such as Hu et al. \cite{Hu2009} and Li et al. \cite{huakang14}. The use of these knowledge bases can help enrich the content inherent to the web documents. In these cases, the classification structure of articles within Wikipedia's taxonomy is leveraged to associate web documents with Wikipedia concepts and categories; this process of linking concepts in documents to Wikipedia articles is also known as wikification \cite{cassidy2012analysis}. Then, Wikipedia entries or their n-grams are matched with documents to expand the content of each document with related content. 

While recent years have seen a growing body of research in the use of fuzzy logic to make the most of the document representation for clustering purposes \cite{LinH13,wang2014improved,zhou2015clustering}, the exploitation of the characteristics of HTML documents, which are rich in structure, remains relatively unexplored. One of the most recent approaches making use of fuzzy logic representation for semi-structured documents is that introduced by Ensan and Biletskiy \cite{ensanB13}. The caveat of this approach is the need of a human in the loop for generating templates, which boosts the system's performance by extracting additional information within a supervised approach. The authors did not however explore an alternative solution for fully automating the process. Our work intends to fill this gap, performing a comprehensive study on the use of the HTML structure and content with fuzzy logic for web document clustering in an unsupervised approach.

The works which are closest to ours are by Molinari and Pasi \cite{molinari96}, focused on an Information Retrieval task, and by Fresno and Ribeiro \cite{Fresno04}, who presented an Analytical Combination of Criteria (ACC) to represent web pages in web page classification and clustering tasks. It is based on a linear combination of different heuristic criteria within the Vector Space Model. These criteria were selected taking into account how a human reader skims through a document to identify the most representative words. The criteria used by ACC are \texttt{title}, \texttt{emphasis}, \texttt{position}, and \texttt{frequency}. Based on the same criteria, Fresno \cite{Fresno06} proposed an approach called Fuzzy Combination of Criteria (FCC), an alternative way of combining them in a non-linear way. In this case, a fuzzy logic based system is employed to define the expert knowledge about how to combine these criteria. The output is also a single vector within the Vector Space Model, representing the estimated importance of each term in a given document. One of the main advantages of FCC is its flexibility, which can be easily utilized for different purposes within different tasks. In fact, recent works have adapted FCC for different purposes, including Nassem et al. \cite{naseem2013near} for the detection of near duplicate web pages, and Bartik \cite{bartik10} for web page classification. The use of fuzzy logic for feature selection and web representation is still an active topic of interest, and is used as can be seen in recent research \cite{kraft2015fuzzy,kolonin2015automatic}. To the best of our knowledge, however, no alternatives to FCC have been proposed, and therefore FCC represents, at the time of this writing, the state of the art in the fully automated, unsupervised fuzzy model for web page representation based on web page structure.

In the present work, we rely on the FCC fuzzy representation as a starting point for our research in order to study the fuzzy combination model in different ways, from analyzing its original definition, to proposing new ways of exploiting the system to perform the combination, as well as to explore the possibility of adapting the system to the input we want to represent. In what follows we further describe the FCC approach, which our work builds on.

\section{FCC: Fuzzy Combinations of Criteria}
\label{sec:fcc}

\begin{figure*}[htb]
  \centering
  \subfloat[Frequency sets]{\label{sf:frequency-sets}\includegraphics[width=100px]{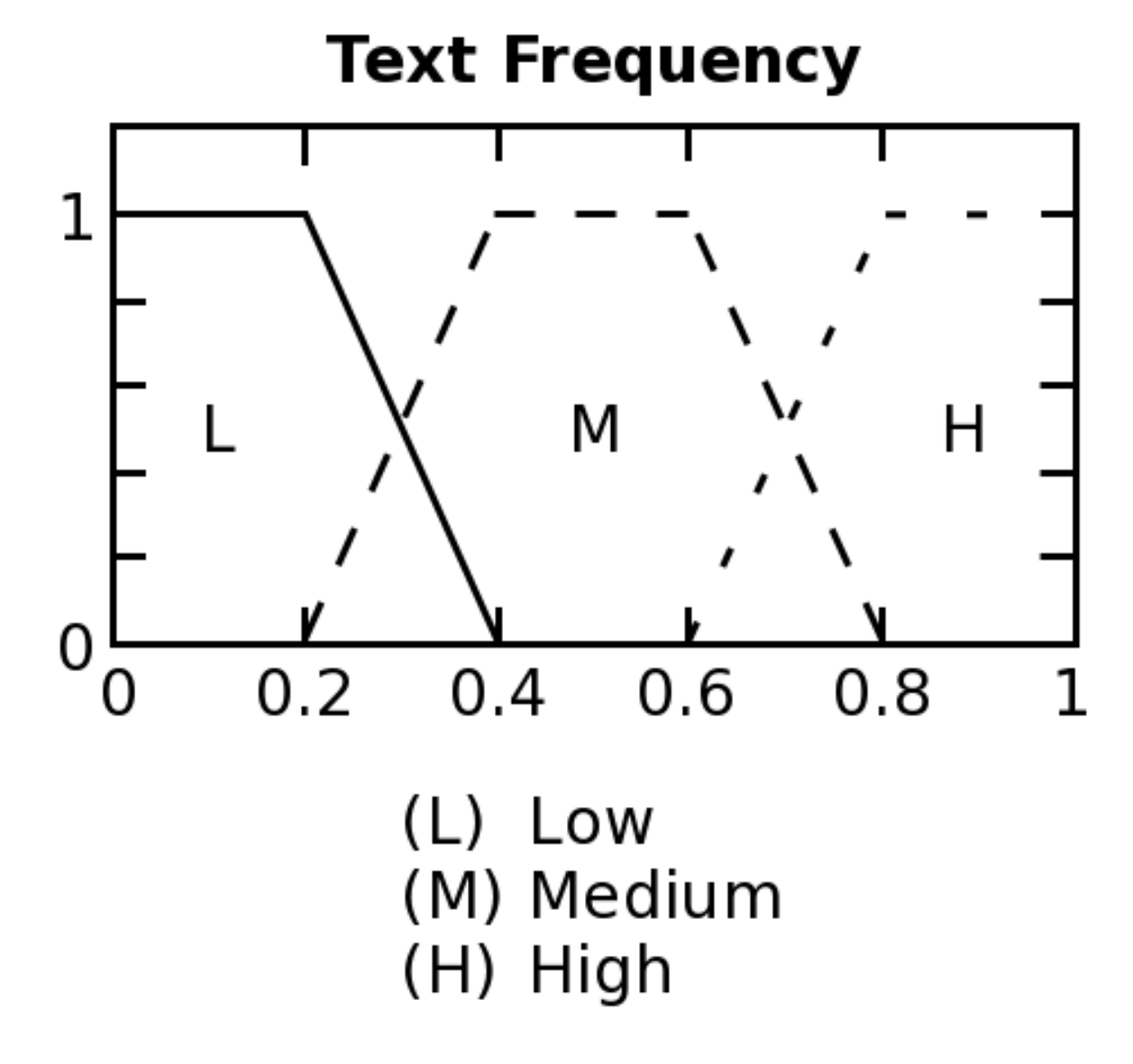}} 
  \hspace{10px}
  \subfloat[Title sets]{\label{sf:title-sets}\includegraphics[width=100px]{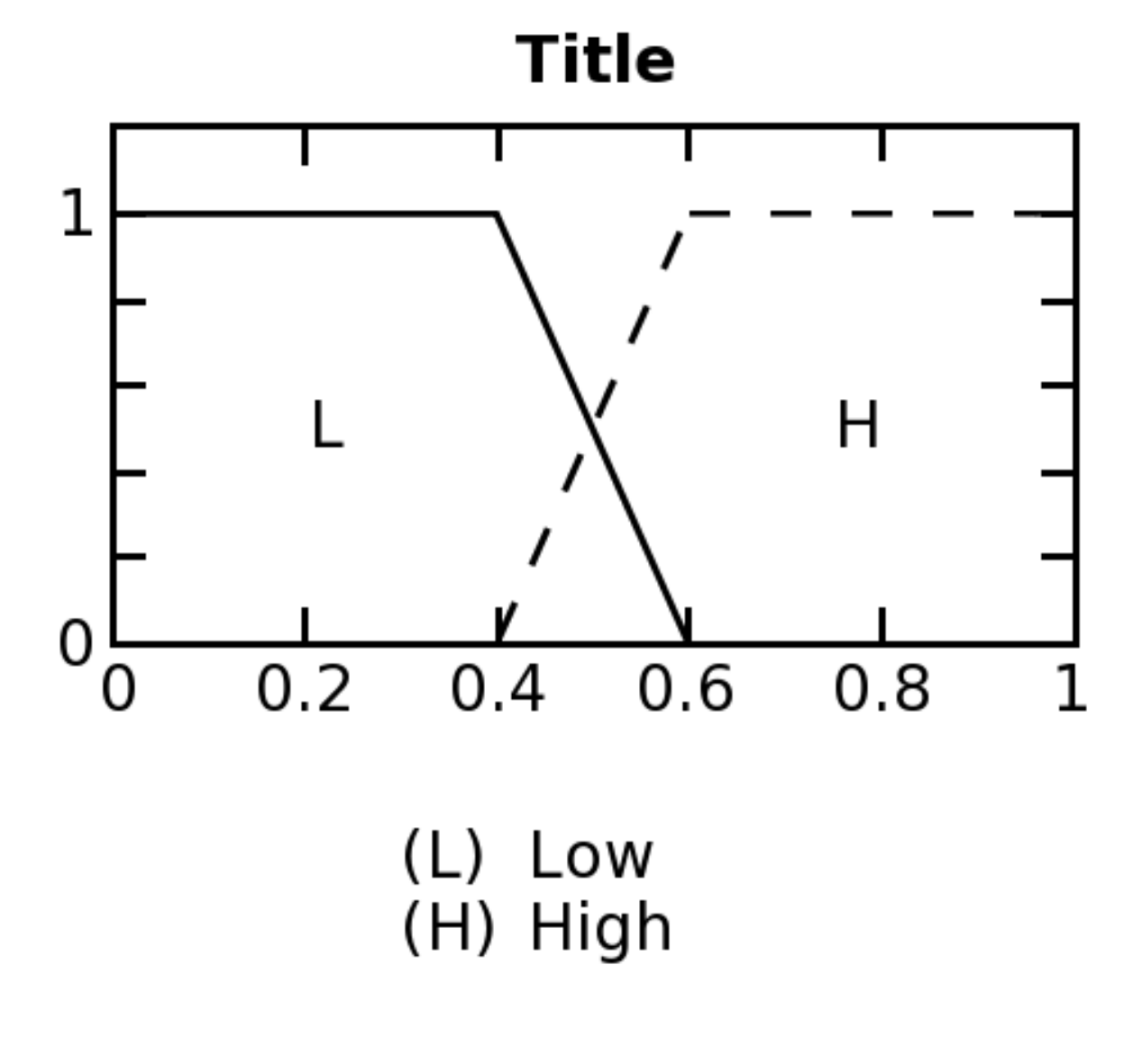}} 
  \hspace{10px}
  \subfloat[Emphasis sets]{\label{sf:emphasis-sets}\includegraphics[width=100px]{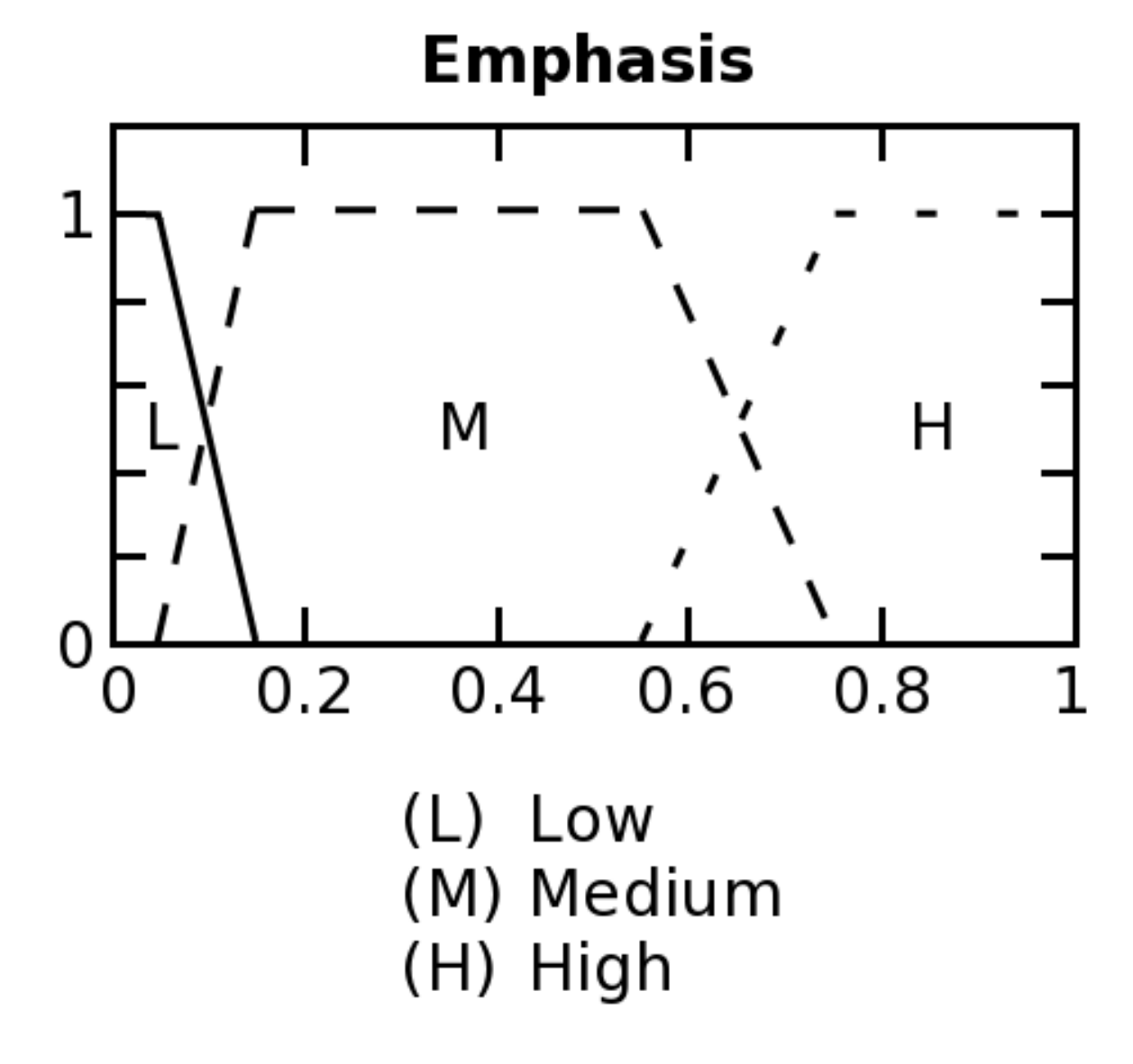}}  
  \hspace{10px}
  \subfloat[Position sets]{\label{sf:position-sets}\includegraphics[width=100px]{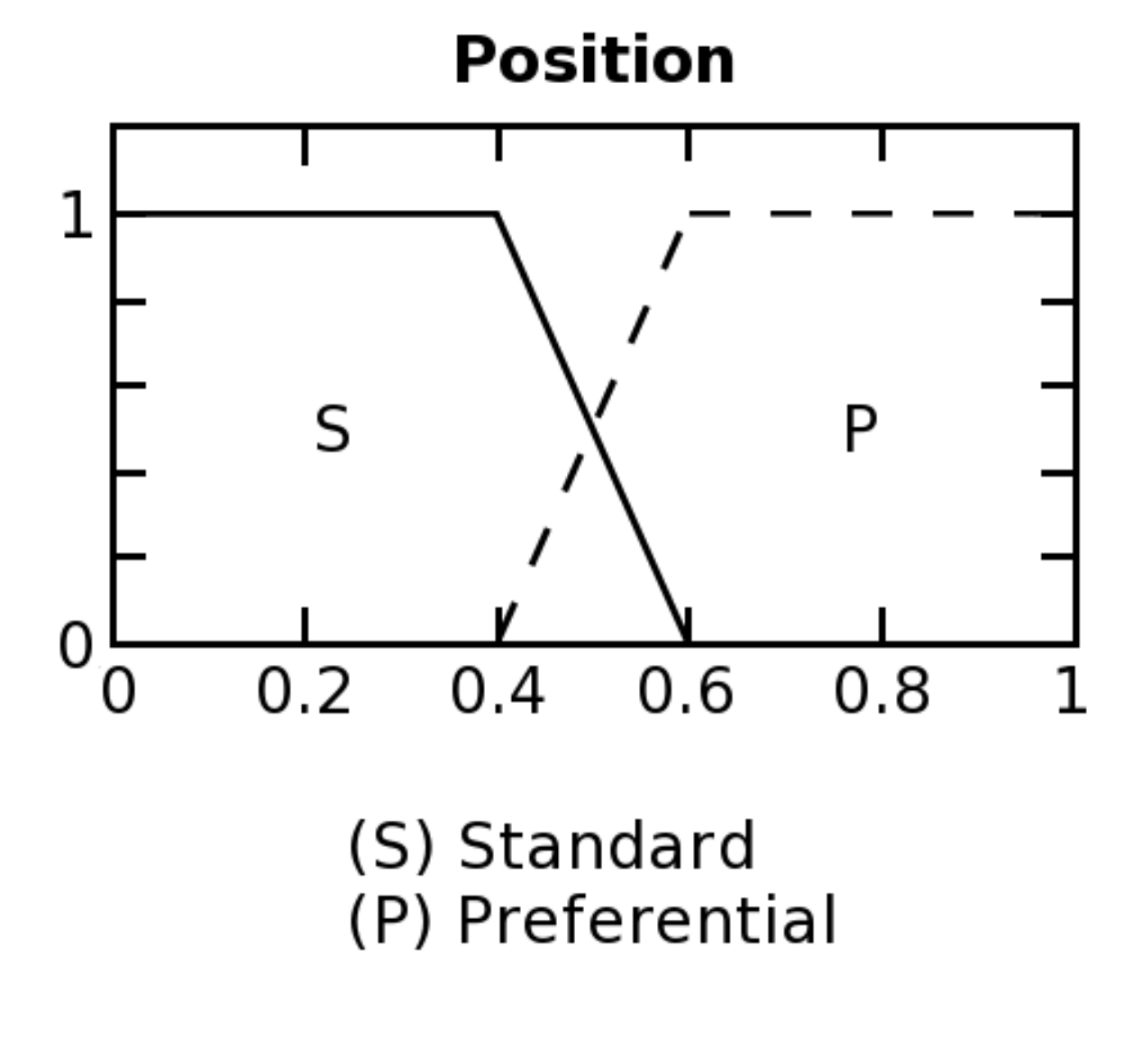}}
  \caption{Data base for FCC. Input linguistic variables.}
  \label{f:fcc-data-base}
\end{figure*}

The fuzzy system in FCC is built over the concept of linguistic variable and its fuzzy sets. Each variable describes the membership degree of an object to a particular class and it is defined by human experts. This membership degree is defined by a membership function. For each heuristic criterion (\texttt{frequency}, \texttt{title}, \texttt{emphasis}, and \texttt{position}), an associated linguistic variable is defined, as well as for the system output (importance):

\begin{enumerate}
\item \textbf{Text Frequency}: term frequency in the document. Its input is calculated by normalizing this frequency to the maximum number of occurrences of any term in that document. It is defined in three fuzzy sets: \emph{low}, \emph{medium}, and \emph{high} (see Figure \ref{sf:frequency-sets}).

\item \textbf{Title}: term frequency within the \texttt{<title>} tag. Its input is calculated by normalizing this frequency to the maximum number of occurrences of any term in the title of that document. It is defined in two fuzzy sets: \emph{low} and \emph{high} (see Figure \ref{sf:title-sets}).

\item \textbf{Emphasis}: term frequency in emphasized parts of the text\footnote{We use a manually created list of HTML tags that add emphasis: \texttt{<em>}, \texttt{<b>}, \texttt{<u>}, \texttt{<strong>}, \texttt{<big>}, \texttt{<h*>}, \texttt{<cite>}, \texttt{<dfn>}, \texttt{<i>}, \texttt{<blockquote>}}. Its input is calculated by normalizing this frequency to the maximum number of occurrences of any term in emphasized text segments in that document. It is composed of three fuzzy sets: \emph{low}, \emph{medium} and \emph{high} (see Figure \ref{sf:emphasis-sets}).

\item \textbf{Position}: the global position of a term in the document, defined in two fuzzy sets: \emph{standard} and \emph{preferential} (see Figure \ref{sf:position-sets}). It is obtained by means of an Auxiliary Fuzzy System that takes as input all the positions of a term within a document (captured by the other linguistic variable \textbf{term position}) and returns the global position value in terms of two fuzzy sets, \emph{standard} and \emph{preferential}. 

\item \textbf{Importance}: it is the output of the fuzzy system and equates to the estimated importance of a term in the document content. It has five homogeneously distributed fuzzy sets: \emph{no}, \emph{low}, \emph{medium}, \emph{high} and \emph{very high}. 
\end{enumerate}

These membership functions have a trapezoidal shape. All the variables except \texttt{emphasis} are defined by sets of equal size symmetrically distributed along the possible input values. These sets were defined without restricting to specific datasets. However, \texttt{emphasis} is considered separately because when the maximum frequency value for emphasized words in a document is small, the normalization could have high impact on the importance of other emphasized terms. For example, using symmetrical sets and having a maximum of $4$ would lead to consider the importance of terms emphasized once as \emph{low}, when we may want to increase the importance of these terms. For this reason, the sets for \texttt{emphasis} were asymmetrically defined. This way, frequencies that would be strictly \emph{low} can also be considered as \emph{medium}, since we can expect small maximum values in \texttt{emphasis}.

The other part of the knowledge base is a set of IF-THEN rules. The aim of the rules is to combine one or more input fuzzy sets (antecedents or premises) and to associate them with an output fuzzy set (consequent). Once the consequents of each rule have been calculated, and after an aggregation stage, the final set is obtained representing the word based on its importance within the document content. The complete set of 31 rules defined in the FCC approach can be found in \cite[p. 130]{Fresno06}. Example \ref{ex:ifthen1} shows an example of an IF-THEN rule.\\

\begin{xmpl}
\label{ex:ifthen1}

\textit{IF Title IS High AND Frequency IS Low AND Emphasis IS Low AND Position is Standard THEN Importance IS Low}\\

\end{xmpl}

%

The rule set is complete, so that every possible input has to trigger at least one rule. The inference engine evaluates all the triggered rules on the basis of the \emph{Center Of Mass} (COM) algorithm, which weighs the output of every triggered rule, taking into account the truth degree of their antecedents. It takes the balance point or centroid of all the scaled membership functions taken together for that variable \cite{hopgood2011}. The output for each term input to the system is calculated by scaling the membership functions by product and combining them by summation. 

The rule base presented in \cite{Fresno06} relies on the following three considerations: 

\begin{enumerate}
\item If a word appears in the title or the word is emphasized, it should also appear in one of the other criteria in order to be considered important. This aims to alleviate the problem of rhetoric titles or non-informative highlighting; 

\item Words occurring in the beginning or in the end of a document are more likely to be important than the rest of the words, as some documents contain overviews and summaries in order to attract the interest of the reader. When the words in a preferential position do not occur also in the title or emphasized, then we could assume that the document does not adhere to the mentioned structure and we could reduce the importance value of that word; 

\item It might be the case that there are no emphasized words in a document, the document has no title, or the title has no important words. In these cases we have to take care of the penalization it could cause to the combination. If the previous criteria did not pick important words, the word frequencies in the whole document are used. Different from the others, the \texttt{frequency} criterion is always available. 
\end{enumerate}

\section{Experimental Framework}
\label{sec:experimental-framework}

In this section, we describe the experimental settings that we use in our research.

\subsection{Datasets}

To make results comparable to those by Fresno \cite{Fresno06}, we also use the same two datasets, Banksearch \cite{Sinka05} and WebKB \cite{Craven2000}. Additionally, we use the Social-ODP-2k9 Dataset \cite{Zubiaga09b}, which provides the features we need for the extended analysis looking at anchor texts.

\begin{enumerate}
\item \textbf{Banksearch \cite{Sinka05}.} A benchmark dataset designed for evaluation of web page clustering. We use the $10$ main categories --\textit{A} to \textit{J}--,(Commercial Banks, Building Societies, Insurance Agencies, Java, C/C++, Visual Basic, Astronomy, Biology, Soccer, Motor Sports). We removed the other category (\textit{K}, Sport) for being of a different granularity level and hence not comparable to the rest. This results in $9,897$ documents evenly distributed across categories.

\item \textbf{WebKB \cite{Craven2000}.} A dataset that includes web pages from computer science departments of various universities. We use $4,518$ web pages that are categorized into $6$ imbalanced categories (\emph{Student}, \emph{Faculty}, \emph{Staff}, \emph{Department}, \emph{Course}, \emph{Project}), after removing the \emph{Other} miscellanea category that is not comparable to the rest. This dataset is more heterogenous than the others, as web pages on a common subject can be found in different categories, such as \textit{Java programming} categorized into \textit{Student}, \textit{Course} or \textit{Department}.

\item \textbf{Social ODP 2k9 \cite{Zubiaga09b}.} A dataset that consists of HTML documents retrieved from links bookmarked by users on Delicious.com. The classification of these documents was inferred from the taxonomy of the Open Directory Project\footnote{http://www.dmoz.org/}. From this dataset, we used $12,148$ documents that passed a valid HTML test. The documents are classified into $17$ categories. This dataset is also imbalanced, where the most prominent category accounts for 26\% of the documents. In addition to the documents themselves, we collected up to $300$ anchor texts per document in the collection. The anchor texts were retrieved by querying Google for links pointing to collection pages.
\end{enumerate}

\subsection{Baseline}

As a baseline, we compute the weight of each word occurring in a document by using the well-known TF-IDF term weighing function, where the term frequency (TF) in a document is combined with the Inverse Document Frequency (IDF) of that term in the whole collection:

\begin{equation} 
 \mathrm{TF\mbox{-}IDF}(t_i,d_j,D) = \mathrm{TF}(t_i,d_j) \times \log \frac{|D|}{|\{d_j \in D: t_i \in d_j\}|} 
 \label{eq:tfidf}
\end{equation}

\noindent where $t_i$ is a term, $_j$ a document, $D$ the whole corpus, $|D|$ is the total number of documents in the corpus and $|\{d_j \in D: t_i \in d_j\}|$ is the number of documents where the term $t_i$ appears.

\subsection{Dimensionality Reduction}

Dimensionality reduction aims to reduce the number of vector components, consequently attempting to reduce the computational cost while the performance loss is as little as possible. Many different dimensionality reduction approaches have been introduced in the literature, aiming to address the limitations of traditional techniques such as Principal Component Analysis and classical scaling. These approaches range from simpler techniques relying solely on term frequencies, to more complex methods derived from approaches originally defined for text classification. Van der Maaten et al. \cite{Maaten08} present a review and comparison of nonlinear dimensionality reduction techniques, which they group into two types: (1) convex techniques (full spectral or sparse spectral), optimizing an objective function that does not contain any local optima; and (2) non-convex techniques (weighted euclidean distances, alignment of local linear models, or neural networks) that optimize objective functions that do contain local optimal.
 
 
On the other hand, from the perspective of availability and use of labeled data for training, feature selection can be categorized as supervised, semisupervised or unsupervised. When it comes to supervised approaches, He et. al \cite{he2005} introduce a feature selection algorithm called Laplacian Score, and Kala et al. \cite{kala2009} use Fuzzy C Means clustering to find clusters in the given training data set. Others like \cite{xu2012} and \cite{wanga15} introduced approaches within a semi-supervised learning scenario. For an unsupervised scenario, in the abscense of class information, there are feature selection and dimensionality reduction methods which preserve the local geometrical structure such as Multi-Cluster Feature Selection \cite{cai2010} and L1 Graph Based on Sparse Coding for Feature Selection \cite{Xu2013}.

We introduce a unsupervised reduction method called Most Frequent Terms (MFT), which is based on term importance estimated by a term weighing function. The MFT method works as follows. First, the terms in each document are ranked based on the values of the weighing function. Then, the terms in the first position of the ranked list of each document are sorted according to the number of times they occur in the rankings. In case of a tie, we order them according to the maximum weight between them. We then do the same for terms in the second position of ranking, in the third position, and so on. The process stops when the desired number of terms is reached.
Even though the resulting list may be larger than the size sought, the ordered list enables us to get the exact number of terms from the top. 

As an alternative dimensionality reduction method, we also compare Latent Semantic Indexing (LSI) \cite{Landauer98}. LSI projects the initial space of documents and their words into a reduced vector space, where the mapping is performed in such a way that the independence is kept for terms that do not co-occur.

\subsection{Clustering Algorithm}

We chose Cluto rbr (k-way repeated bisections globally optimized) as the clustering algorithm \cite{karypis03} for our experiments. The number of clusters $k$ is set to the number of categories in each dataset to make the evaluation process more intuitive. Having $k$ set to the actual number of clusters enables to explore differences between representation approaches, leaving aside the effect of the selection of the number of clusters. The rest of the parameters are set to their default values.

\subsection{Evaluation Measure}

We use the $F_1$ score \cite{Rijsbergen1974} as the evaluation measure (see Equation \ref{eq:f-measure}).

\begin{equation}
 F_1 = 2 \cdot \frac{precision \cdot recall}{precision + recall}
 \label{eq:f-measure}
\end{equation}

where precision and recall are defined as follows:

\begin{equation}
 precision=\frac{|\{relevant\ docs\}\cap\{retrieved\ docs\}|}{|\{retrieved\ docs\}|}
 \label{eq:precision}
\end{equation}

\begin{equation}
 recall=\frac{|\{relevant\ docs\}\cap\{retrieved\ docs\}|}{|\{relevant\ docs\}|}
 \label{eq:recall}
\end{equation}

From these, the $F_1$ score for each category can be computed. The overall $F_1$ score is computed as the weighted average of the $F_1$ scores for each category.

\section{Improving the Combination of Criteria}
\label{sec:improving-combination}

In this section, we evaluate the performance of the state-of-the-art fuzzy logic approach FCC. We also propose and evaluate two novel alternative approaches, EFCC and AddFCC.

\subsection{Study of FCC and Individual Criteria}

As an initial comparative study over the existing FCC approach, we propose and analyze four variations of this term weighing function, one for each criterion, in such a way that the output of the system will depend only on one criterion at a time. Table \ref{rules:emphfcc} shows an example rule base for a system that solely relies on the \texttt{emphasis} criterion to determine the output.

\begin{table*}[htbp]
\begin{center}
\footnotesize
\caption{Rule base for the system based on \texttt{emphasis} criterion.}
\label{rules:emphfcc}
\begin{tabular}[c]{p{1mm}p{4mm}p{5mm}p{12mm}p{5mm}p{11mm}p{5mm}p{13mm}p{7mm}p{16mm}}
\hline
{\bf IF} & {\bf Title} & {\bf AND} & {\bf Frequency} & {\bf AND} & {\bf Emphasis} & {\bf AND} & {\bf Position} & {\bf THEN} & {\bf Importance} \\ 
\hline
  & & & & & High & & & ~~~$\Rightarrow$  & Very High \\ 
  & & & & & Medium & & & ~~~$\Rightarrow$  & Medium \\ 
  & & & & & Low & & & ~~~$\Rightarrow$  & No \\ 
\hline \\
\end{tabular}
\end{center}
\end{table*}

We used the MFT reduction given that it selects the highest weighted features without transforming them. This enables us to perform a fairer comparison of different term weighing approaches.

\begin{figure}
 \caption{Graphical representation of data in Table \ref{table:critan}}

 \includegraphics[width=\textwidth]{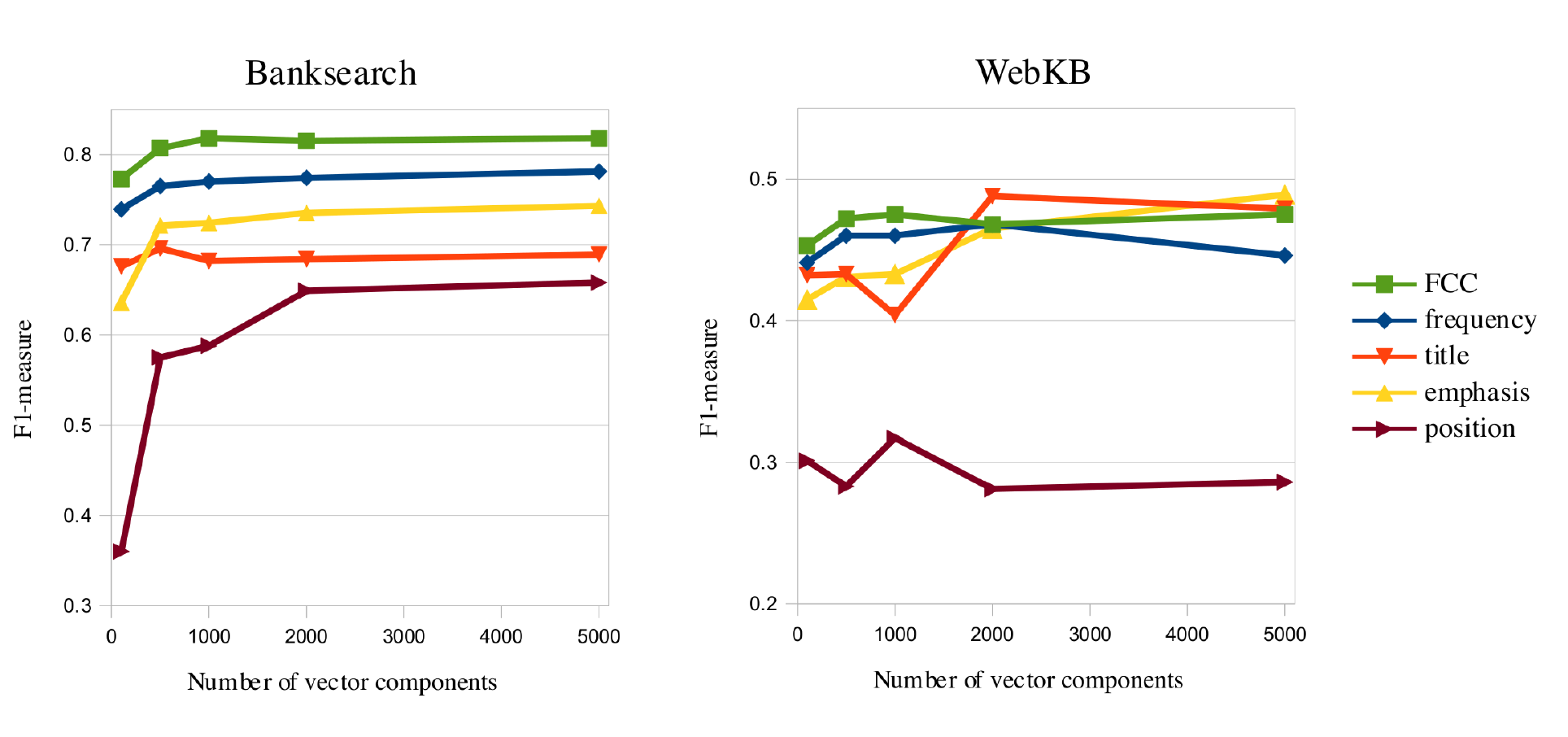}
 \label{fig:critan}

 \footnotesize

 \centering
 \begin{tabular}{| l | c | c | c | c | c | c |}
   \hline
   \textbf{Rep.} & \textbf{100} & \textbf{500} & \textbf{1,000} & \textbf{2,000} & \textbf{5,000} & \textbf{Avg.} \\ 
   \hline 
   \multicolumn{7}{| c |}{\textbf{Banksearch}} \\
   \hline
   FCC & \textbf{0.723} & \textbf{0.757} & \textbf{0.768} & \textbf{0.765} & \textbf{0.768} & \textbf{0.756} \\
   title & 0.626 & 0.646 & 0.632 & 0.634 & 0.639 & 0.635 \\ 
   emphasis & 0.586 & 0.671 & 0.674 & 0.685 & 0.693 & 0.662 \\ 
   frequency & 0.689 & 0.715 & 0.720 & 0.724 & 0.731 & 0.716 \\ 
   position & 0.310 & 0.525 & 0.538 & 0.599 & 0.608 & 0.516 \\ 
   \hline
   \multicolumn{7}{| c |}{\textbf{WebKB}} \\
   \hline 
   FCC & \textbf{0.453} & \textbf{0.472} & \textbf{0.475} & 0.468 & 0.475 & \textbf{0.469} \\ 
   title & 0.432 & 0.433 & 0.404 & \textbf{0.488} & 0.479 & 0.447 \\ 
   emphasis & 0.415 & 0.431 & 0.433 & 0.465 & \textbf{0.489} & 0.447 \\ 
   frequency & 0.441 & 0.460 & 0.460 & 0.468 & 0.446 & 0.455 \\
   position & 0.301 & 0.283 & 0.317 & 0.281 & 0.286 & 0.294 \\ 
   \hline
 \end{tabular}
 \captionof{table}{F1 results for criteria analysis experiments (all with MFT reduction)}
 \label{table:critan}

\end{figure}

Table \ref{table:critan} and Figure \ref{fig:critan} show the results of each individual criterion compared to FCC, where each column shows the performance for different numbers of features ranging from 100 to 5,000, as well as the average. 

For Banksearch, FCC outperforms all individual alternatives, showing the importance of the combination of criteria. Among the individual criteria, \texttt{frequency} performs best, while \texttt{position} is the worst. 

The results for WebKB are quite different. On one hand, \texttt{frequency} is not always the best among individual criteria and, on the other hand, FCC does not always outperform individual criteria, specifically \texttt{title} and \texttt{emphasis} obtain equal or higher F1-measure values in some cases when the vector dimensions are reduced to $2,000$ and $5,000$, respectively. 

In this collection, the frequency distribution of emphasized terms shows a more restricted use of emphasis. It could be due to the limited number of web domains and the similarity among web page contents that only come from Universities. These factors could limit the number of different writing styles, fact that would be reflected in a less scattered distribution of emphasized term frequencies. The same consideration about the restrictions on the creation of WebKB can explain the good results achieved by the \texttt{title} criterion. We can expect that authors use titles in a similar way to emphasis, as both resources are used to highlight important words. In the cases where \texttt{title} and \texttt{emphasis} lead to a better clustering, their combination with \texttt{frequency} and \texttt{position} harms the results. In particular, WebKB documents within categories can be much more heterogeneous than in Banksearch, fact that negatively affects the \texttt{frequency} criterion; the combination should help correct this issue, but it does not. Thus, it suggests that \texttt{frequency} and \texttt{position} are hindering the combination.

\subsection{AddFCC and EFCC: Modifying the Knowledge Base}

The first step to try to improve the fuzzy combination is to understand the bad performance of FCC in WebKB. In the rules of FCC \cite{Fresno06}, when \texttt{frequency} is \emph{low}, the output can be \emph{very high} (the maximum) depending on \texttt{position}, if \texttt{title} and \texttt{emphasis} are \emph{high}. As we saw before, \texttt{frequency} contributes to a good clustering much more than \texttt{position}, so the output should reflect that fact. But, in this case, \texttt{frequency} is totally ignored. This occurs again when \texttt{title} is \emph{low} and \texttt{frequency} \emph{medium}. Both criteria are important for a good grouping, but the output is \emph{very high} based on \texttt{position}, the same as the previous case. In these cases we are underestimating the discrimination power of \texttt{frequency} and \texttt{title}. The same happens when \texttt{frequency} is \emph{medium}, being \texttt{title} and \texttt{emphasis} \emph{low}: \texttt{position} decides again that \texttt{importance} can be the minimum or not, but \texttt{frequency} should count more than \texttt{position}.

On the other hand, the whole set of 31 rules in FCC makes the possible combinations more difficult to understand and evaluate. As the fuzzy system is able to combine the conclusions of the rules, an alternative that we propose is the use of a set of single-input rules for each criterion. Thus, the alternative system calculates the output by combining the different outputs of the fired rules. We refer to this approach AddFCC, whose rule base is shown in Table \ref{rules:addfcc}, which reduces the number of cases that are set to the minimum to keep the rule set complete.

\begin{table*}[htbp]
\begin{center}
\footnotesize
\caption{Rule base for AddFCC. Inputs are related to normalized term frequencies.}
\label{rules:addfcc}
\begin{tabular}[c]{p{1mm}p{4mm}p{5mm}p{12mm}p{5mm}p{11mm}p{5mm}p{13mm}p{7mm}p{16mm}}
\hline
{\bf IF} & {\bf Title} & {\bf AND} & {\bf Frequency} & {\bf AND} & {\bf Emphasis} & {\bf AND} & {\bf Position} & {\bf THEN} & {\bf Importance} \\ 
\hline
  & High & & & & & & & ~~~$\Rightarrow$  & Very High\\ 
  & Low  & & & & & & & ~~~$\Rightarrow$  & No  \\ 
\hline
  & & & High & & & & & ~~~$\Rightarrow$  & Very High  \\ 
  & & & Medium & & & & & ~~~$\Rightarrow$  & Medium  \\ 
  & & & Low & & & & & ~~~$\Rightarrow$  & No  \\ 
\hline
  & & & & & High & & & ~~~$\Rightarrow$  & Very High \\ 
  & & & & & Medium & & & ~~~$\Rightarrow$  & Medium \\ 
  & & & & & Low & & & ~~~$\Rightarrow$  & No \\ 
\hline
  & & & & & & & Preferential & ~~~$\Rightarrow$  & Very High  \\ 
  & & & & & & & Standard & ~~~$\Rightarrow$  & No  \\ 
\hline \\
\end{tabular}
\end{center}
\end{table*}

Since the reduced expressiveness of AddFCC system may give rise to mistakes due to a bad specification of the heuristic knowledge, we introduce another intermediate approach, Extended Fuzzy Combination of Criteria (EFCC). Its rule base combines some criteria explicitly and for others lets the combination to the fuzzy engine (see Table \ref{rules:efcc}). It has two sets of rules: one for \texttt{frequency} and one for the rest of the criteria. This guarantees having at least one rule of each set fired by the system. This avoids underestimation of \texttt{frequency} while also reducing the discriminative power of \texttt{position}.

\begin{table*}[htbp]
\begin{center}
\footnotesize
\caption{Rule base for EFCC. Inputs are related to normalized term frequencies.}
\label{rules:efcc}
\begin{tabular}[c]{p{1mm}p{4mm}p{5mm}p{12mm}p{5mm}p{11mm}p{5mm}p{13mm}p{7mm}p{16mm}}
\hline
{\bf IF} & {\bf Title} & {\bf AND} & {\bf Frequency} & {\bf AND} & {\bf Emphasis} & {\bf AND} & {\bf Position} & {\bf THEN} & {\bf Importance} \\ 
\hline 
  & High &   &   &   & High &   &   & ~~~$\Rightarrow$  & Very High\\ 
\hline
  & High &   &   &   & Medium &   & Preferential & ~~~$\Rightarrow$  & High  \\ 
  & High &   &   &   & Medium &   & Standard & ~~~$\Rightarrow$  & Medium  \\ 

  & High &   &   &   & Low &   & Preferential & ~~~$\Rightarrow$  & Medium  \\ 
  & High &   &   &   & Low &   & Standard & ~~~$\Rightarrow$  & Low  \\ 
  & Low &   &   &   & High &   & Preferential & ~~~$\Rightarrow$  & High  \\
  & Low &   &   &   & High &   & Standard & ~~~$\Rightarrow$  & Medium  \\

  & Low &   &   &   & Medium &   & Preferential & ~~~$\Rightarrow$  & Medium  \\
  & Low &   &   &   & Medium &   & Standard & ~~~$\Rightarrow$  & Low  \\

  & Low &   &   &   & Low &   & Preferential & ~~~$\Rightarrow$  & Low  \\
  & Low &   &   &   & Low &   & Standard & ~~~$\Rightarrow$  & No \\
\hline
  &   &   & High &   &   &   &   & ~~~$\Rightarrow$  & Very High\\ 
  &   &   & Medium &   &   &   &   & ~~~$\Rightarrow$  & Medium  \\ 
  &   &   & Low &   &   &   &   & ~~~$\Rightarrow$  & No \\ 
\hline \\
\end{tabular}
\end{center}
\end{table*}

\begin{figure}
 \centering
 \caption{Graphical representation of data in Table \ref{table:efcc1}.}
 \includegraphics[width=\textwidth]{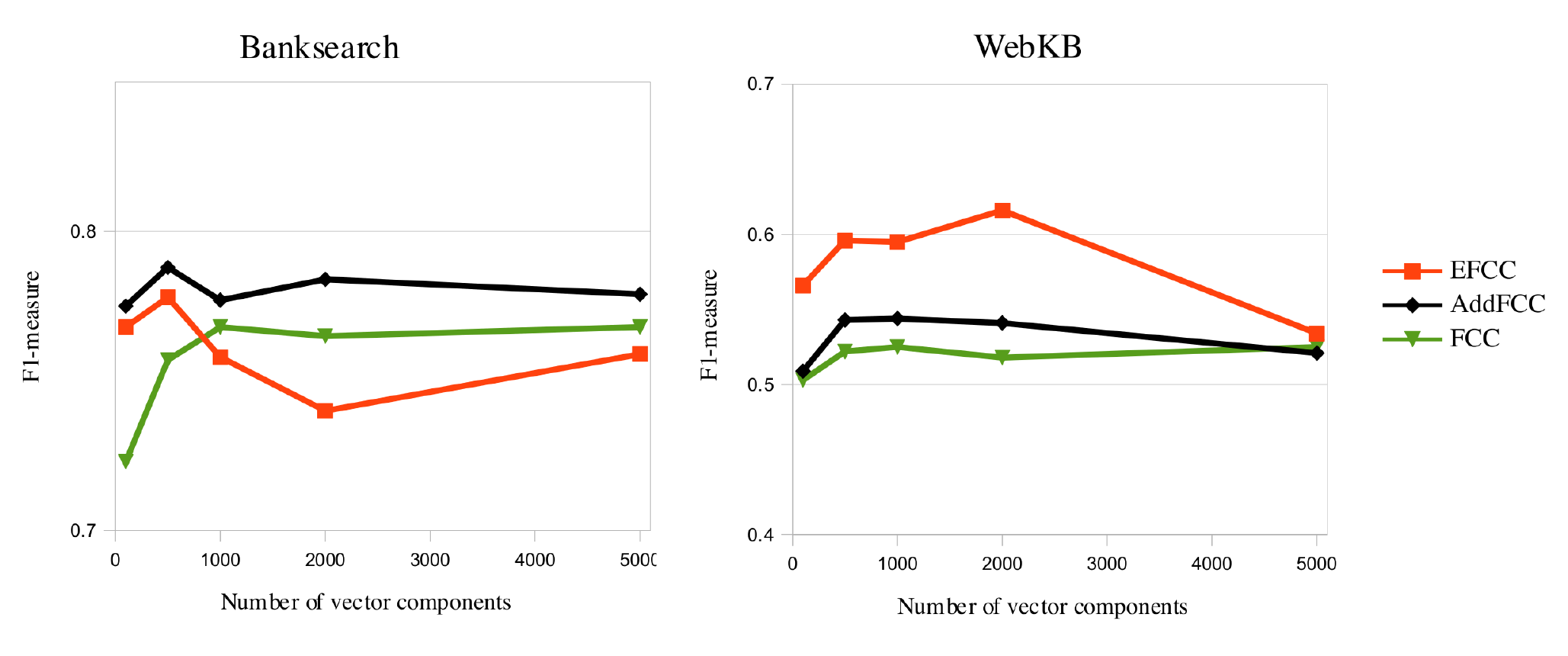}
 \label{fig:efcc1}
 
 \footnotesize
 \begin{tabular}{| l | r | r | r | r | r | r |}
  \hline
  \textbf{Rep.} & \textbf{100} & \textbf{500} & \textbf{1,000} & \textbf{2,000} & \textbf{5,000} & \textbf{Avg.} \\ 
  \hline 
  \multicolumn{7}{| c |}{\textbf{Banksearch}} \\
  \hline
  FCC & 0.723 & 0.757 & 0.768 & 0.765 & 0.768 & 0.756 \\
  EFCC & 0.768 & 0.778 & 0.758 & 0.740 & 0.759 & 0.760 \\
  AddFCC & \textbf{0.775} & \textbf{0.788} & \textbf{0.777} & \textbf{0.784} & \textbf{0.779} & \textbf{0.781} \\ 
  \hline
  \multicolumn{7}{| c |}{\textbf{WebKB}} \\ 
  \hline
  FCC & 0.453 & 0.472 & 0.475 & 0.468 & 0.475 & 0.469 \\
  EFCC & \textbf{0.516} & \textbf{0.546} & \textbf{0.545} & \textbf{0.566} & \textbf{0.484} & \textbf{0.532} \\
  AddFCC & 0.459 & 0.493 & 0.494 & 0.491 & 0.471 & 0.482 \\
  \hline
 \end{tabular}
 \captionof{table}{Fuzzy logic-based alternatives in terms of F1 (all with MFT reduction).}
 \label{table:efcc1}
\end{figure}

Table \ref{table:efcc1} and Figure \ref{fig:efcc1} show the clustering results for AddFCC and EFCC, which are compared to FCC. We observe that EFCC improves FCC clustering results in WebKB in all cases while AddFCC does not, while AddFCC outperforms the other approaches for Banksearch in all cases. Nevertheless, EFCC also achieves good results in Banksearch, particularly with small feature sets. AddFCC has the problem of considering all criteria equally important, and hence overestimating \texttt{position} in the combination, as we observed with FCC too.

\begin{figure}
 \centering
 \caption{Graphical representation of data in Table \ref{table:efcc2}.}
 \includegraphics[width=\textwidth]{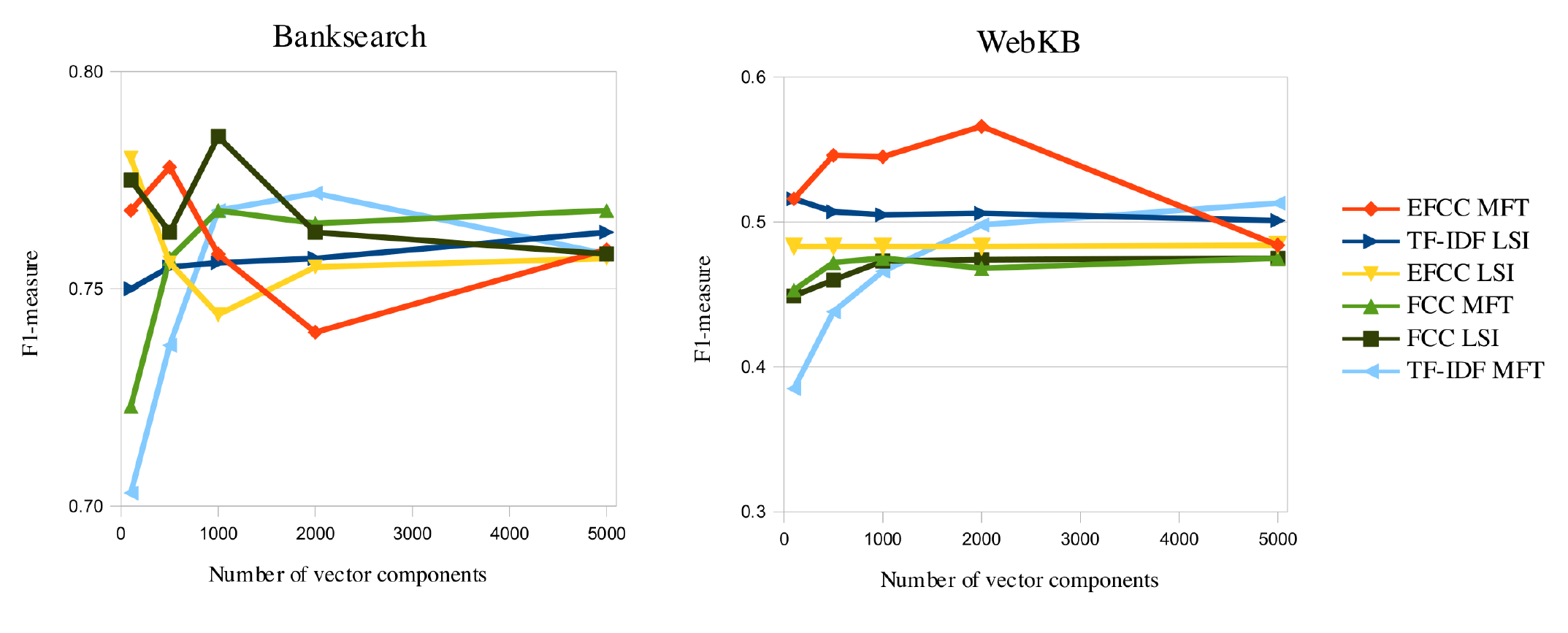}
 \label{fig:efcc2}

 \footnotesize
 \begin{tabular}{| l | r | r | r | r | r | r |}
  \hline
  \textbf{Rep.$\backslash$Dim.} & \textbf{100} & \textbf{500} & \textbf{1,000} & \textbf{2,000} & \textbf{5,000} & \textbf{Avg.} \\ 
  \hline 
  \multicolumn{7}{| c |}{\textbf{Banksearch}} \\
  \hline
  TF-IDF LSI & 0.750 & 0.755 & 0.756 & 0.757 & 0.763 & 0.756 \\
  TF-IDF MFT & 0.703 & 0.737 & 0.768 & \textbf{0.772} & 0.758 & 0.748 \\ 
  FCC LSI & 0.775 & 0.763 & \textbf{0.785} & 0.763 & 0.758 & \textbf{0.769} \\
  FCC MFT & 0.723 & 0.757 & 0.768 & 0.765 & \textbf{0.768} & 0.756 \\
  EFCC LSI & \textbf{0.780} & 0.756 & 0.744 & 0.755 & 0.757 & 0.758 \\
  EFCC MFT & 0.768 & \textbf{0.778} & 0.758 & 0.740 & 0.759 & 0.760 \\
  \hline
  \multicolumn{7}{| c |}{\textbf{WebKB}} \\ 
  \hline 
  TF-IDF LSI & \textbf{0.516} & 0.507 & 0.505 & 0.506 & 0.501 & 0.507 \\
  TF-IDF MFT & 0.385 & 0.438 & 0.466 & 0.498 & \textbf{0.513} & 0.460 \\ 
  FCC LSI & 0.449 & 0.460 & 0.473 & 0.474 & 0.475 & 0.466 \\
  FCC MFT & 0.453 & 0.472 & 0.475 & 0.468 & 0.475 & 0.469 \\
  EFCC MFT & \textbf{0.516} & \textbf{0.546} & \textbf{0.545} & \textbf{0.566} & 0.484 & \textbf{0.532} \\
  EFCC LSI & 0.483 & 0.483 & 0.483 & 0.483 & 0.484 & 0.483 \\
  \hline
 \end{tabular}
 \captionof{table}{F1 performance values for different dimensionality reduction methods with EFCC and other previous alternatives.}
 \label{table:efcc2}
\end{figure}

At this point, we opted for EFCC as an alternative to FCC for our subsequent experiments. We also apply LSI and compare the results of EFCC with TF-IDF and FCC (see Table \ref{table:efcc2} and Figure \ref{fig:efcc2}).

Globally, EFCC MFT achieves the most stable results among collections, and is generally the best approach, with a few exceptions in Banksearch. 
If one is thinking of applying the representation to a new collection, EFCC MFT would be the best option. 
It requires fewer terms to achieve its optimal performance for balanced, homogeneous collections. This posits EFCC MFT as a suitable approach to be applied to new, unseen collections. 
Furthermore, the additive properties of the fuzzy system make it possible to reduce the number of rules needed to specify the knowledge base of EFCC and therefore, the system is easier to understand.

On the other hand, the good behavior of MFT depends on the term weighing function applied before. Because of this, we believe that the use of light dimension reduction techniques is a good alternative, at the price of selecting a proper term weighing function, for the clustering problem to solve.

\subsection{Incorporating Context: Criteria Beyond the Document Itself}

Moving away from the sole use of the document's content itself, now we explore the application of two techniques to improve EFCC with contextual information: (1) Inverse Document Frequency, and (2) anchor texts.
 
\subsubsection{Inverse Document Frequency  (IDF)}
\label{ss:efcc_idf}

With IDF we incorporate information from the whole collection to the representation, which we do by using the product of both:

\begin{equation}
 \mathrm{EFCC\mbox{-}IDF}(t_i,d_j,D) = \mathrm{EFCC}(t_i,d_j) \times  \mathrm{IDF}(t_i,D) 
 \label{eq:efccidf}
\end{equation}

where $t_i$ is a term, $d_j$ a document, and $D$ the whole corpus.

\begin{figure}
 \centering
 \caption{Graphical representation of data in Table \ref{table:efccidf}.}
 \includegraphics[width=\textwidth]{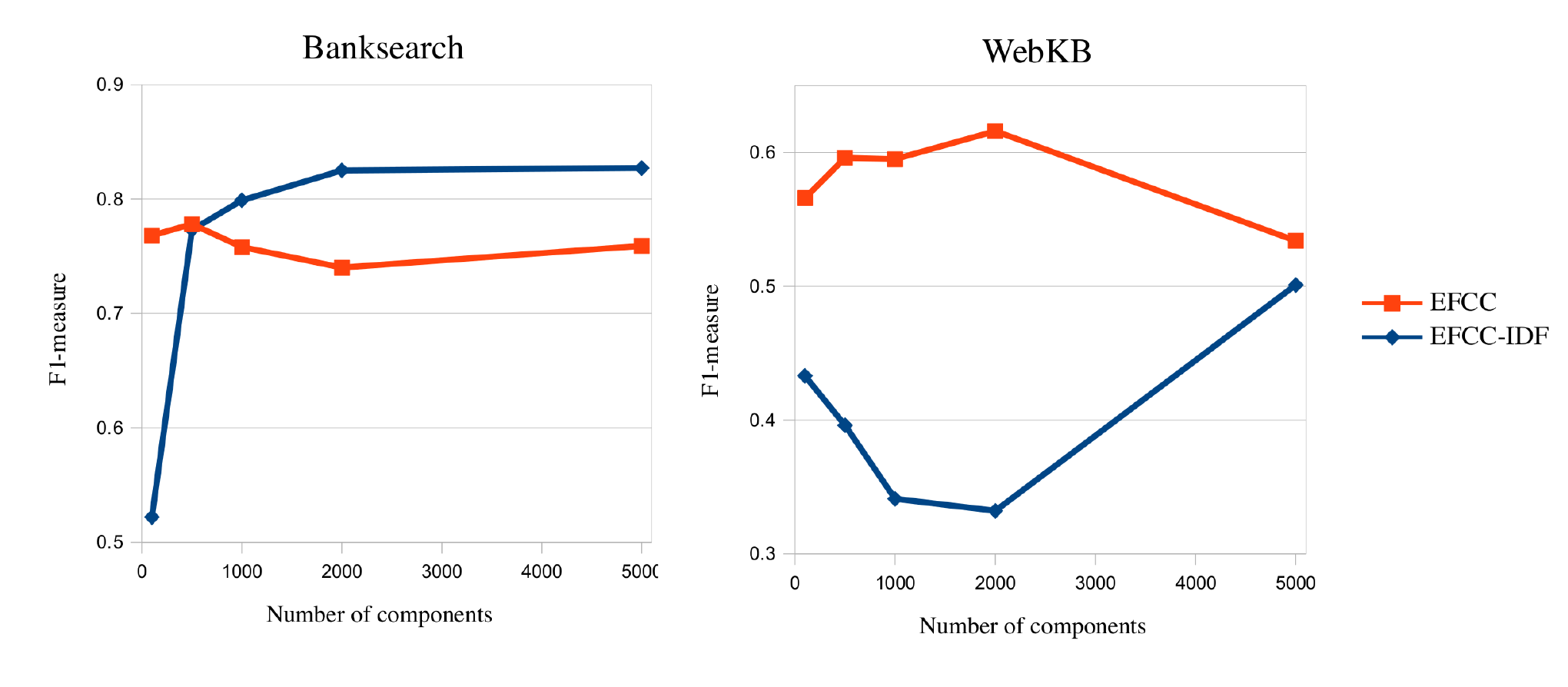}
 \label{fig:efccidf}

 \footnotesize
 \begin{tabular}{| l | r | r | r | r | r | r |}
  \hline
  \textbf{Rep.} & \textbf{100} & \textbf{500} & \textbf{1,000} & \textbf{2,000} & \textbf{5,000} & \textbf{Avg.} \\
  \hline
  \multicolumn{7}{| c |}{\textbf{Banksearch}} \\
  \hline
  EFCC & \textbf{0.768} & \textbf{0.778} & 0.758 & 0.740 & 0.759 & \textbf{0.760} \\
  EFCC-IDF & 0.522 & 0.773 & \textbf{0.799} & \textbf{0.825} & \textbf{0.827} & 0.749 \\
  \hline
  \multicolumn{7}{| c |}{\textbf{WebKB}} \\
  \hline 
  EFCC & \textbf{0.516} & \textbf{0.546} & \textbf{0.545} & \textbf{0.566} & \textbf{0.484} & \textbf{0.532} \\
  EFCC-IDF & 0.383 & 0.346 & 0.291 & 0.282 & 0.451 & 0.350 \\
  \hline
 \end{tabular}
 \captionof{table}{F1 results for $\mathrm{EFCC\mbox{ }IDF}$ experiments (all with the MFT reduction method).}
 \label{table:efccidf}
\end{figure}

Looking at the Table \ref{table:efccidf} and Figure \ref{fig:efccidf}, EFCC-IDF works really well with over $500$ features in Banksearch, but much worse with $100$. WebKB EFCC IDF results are much worse in all cases. This is due to the penalization that IDF applies to common terms. In a clustering task, instead, we look for terms that are common across documents of the same group. Hence, this suggests that the combination of EFCC and IDF is not suitable for the purposes of a clustering task.

%

\subsubsection{Anchor Texts}
\label{ss:efcc_anchor}


There are a number of ways of adding anchor texts to document representation methods. We are interested in elucidating whether anchor texts could help improve web page representation in clustering or not, but at the same time, we want to investigate different alternatives for the combination within a term weighing function. 

To analyze whether and how anchor texts can contribute to the document representation, we explore two different ways of incorporating them using EFCC: (a) Appended to the document's content itself, and hence contributing to the \texttt{frequency} criterion; and (b) Appended to the document's title, and therefore contributing at the same level as the title itself. These approaches considering anchor texts are in line with those described by Wang and Kitsuregawa \cite{Wang2002} and Huang et al. \cite{Huang2006}.


We did the experiments in three different settings in each case: (1) Adding anchor texts; (2) Adding anchor texts and removing textual content from outlinks; and (3) Removing words that are frequently used across anchor texts, such as 'click', 'link' or 'homepage'.

We use the SODP dataset in these experiments, as it is the only dataset that includes anchor texts. As it is a new dataset not explored in previous sections, we also compare results with FCC and AddFCC.

\begin{figure}
 \centering
 \caption{Graphical representation of data in Table \ref{table:anchor}.}
 \includegraphics[width=\textwidth]{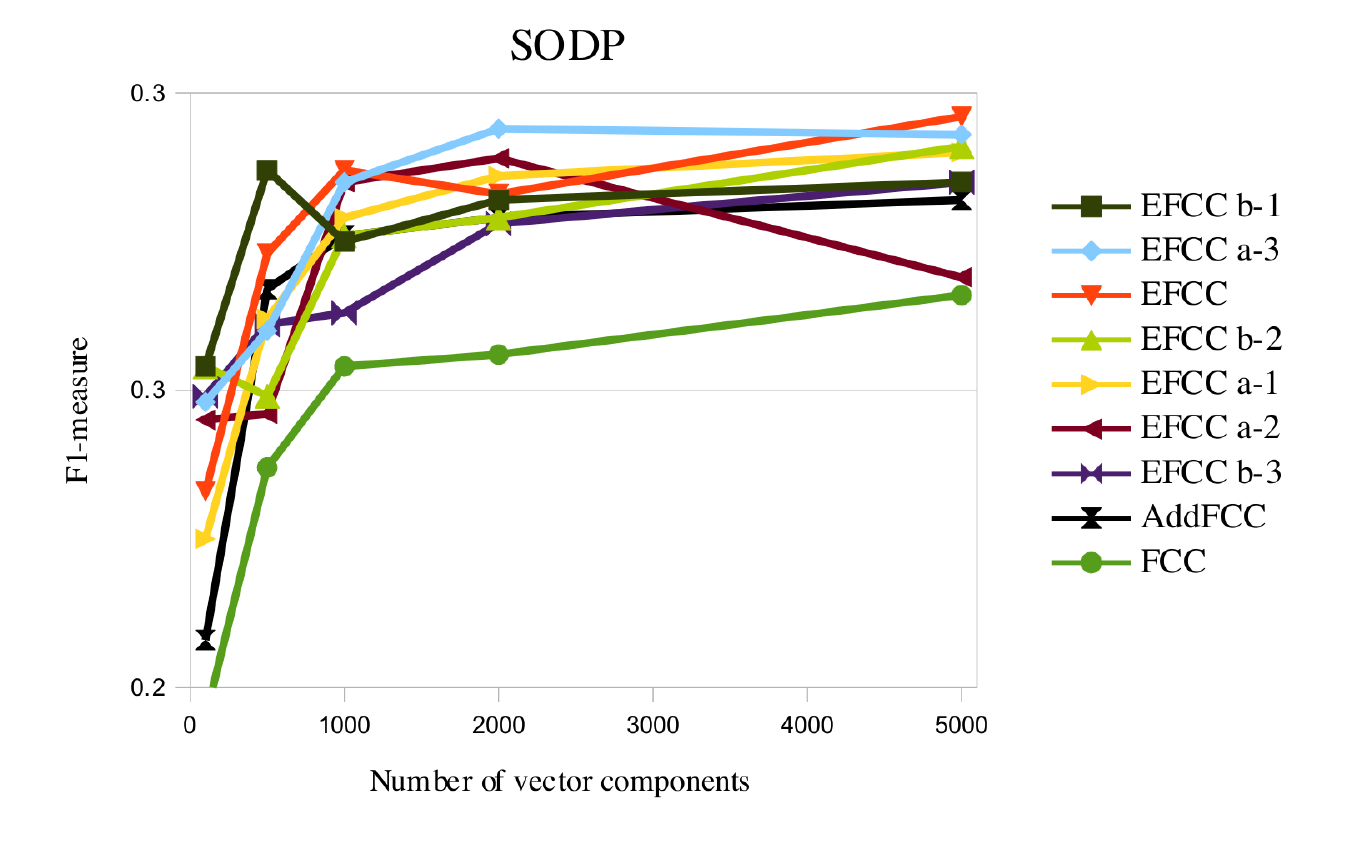}
 \label{fig:anchor}

 \footnotesize
 \begin{tabular}{| l | r | r | r | r | r | r |}
  \hline
  \textbf{Rep.} & \textbf{100} & \textbf{500} & \textbf{1,000} & \textbf{2,000} & \textbf{5,000} & \textbf{Avg.} \\ 
  \hline 
  \multicolumn{7}{| c |}{\textbf{SODP}} \\
  \hline
  FCC & 0.195 & 0.237 & 0.254 & 0.256 & 0.266 & 0.242 \\ 
  AddFCC & 0.208 & 0.267 & 0.276 & 0.279 & 0.282 & 0.262 \\ 
  EFCC & 0.233 & 0.273 & \textbf{0.287} & 0.283 & \textbf{0.296} & 0.275 \\
  EFCC a-1 & 0.225 & 0.262 & 0.279 & 0.286 & 0.290 & 0.268 \\
  EFCC a-2 & 0.245 & 0.246 & 0.285 & 0.289 & 0.269 & 0.267 \\
  EFCC a-3 & 0.248 & 0.260 & 0.285 & \textbf{0.294} & 0.293 & 0.276 \\
  EFCC b-1 & \textbf{0.254} & \textbf{0.287} & 0.275 & 0.282 & 0.285 & \textbf{0.277} \\
  EFCC b-2 & \textbf{0.254} & 0.249 & 0.276 & 0.279 & 0.291 & 0.270 \\
  EFCC b-3 & 0.249 & 0.261 & 0.263 & 0.278 & 0.285 & 0.267 \\ 
  \hline 
 \end{tabular}
 \captionof{table}{F1 results for anchor text experiments (all with MFT reduction).}
 \label{table:anchor}
\end{figure}

Table \ref{table:anchor} and Figure \ref{fig:anchor} show the results of different alternatives using anchor texts. Each approach has a letter and a number appended, referring to the way in which anchor texts are exploited, as described above. 
The first three rows of the table show that EFCC outperforms FCC and AddFCC in all cases. This corroborates the limitations of FCC, and reinforces our motivation looking into an alternative approach where not all the criteria contribute equally to the combination. When it comes to the contribution of anchor texts, no approach improves EFCC clearly in all the cases, as the slight differences suggest when looking at the averages. Anchor texts do have a positive impact when we use vectors of small size, particularly when the terms in the anchor texts are considered as page titles (b alternative). However, as we increase the size of the vectors, anchor texts are not useful any more, leading to worse performance. Regarding the use of anchor texts as titles, the best option is to just add anchor texts as title terms (named b-1). 
The slight improvement achieved with anchor texts might not always pay off, given that the collection of anchor texts is a time consuming process.

Different reasons might explain the unsatisfactory results using anchor texts. The collection may have a link structure that is not sufficiently dense, or anchor texts might not be descriptive enough, hence not enabling to capture the topic of documents. This finding is in line with Eiron and McCurley \cite{Eiron2003} and Noll and Meinel \cite{Noll2008}, where authors posited that anchor text terms rather resemble terms used in search queries.

\section{AFCC: Analyzing and Tuning the Membership Functions}
\label{sec:afcc}

We now set forth a proposal to tune the membership functions, which leads to the definition of a revised and novel approach called Abstract Fuzzy Combination of Criteria (AFCC). We first perform a qualitative analysis of the membership functions that we are utilizing, and then we test AFCC, evaluating and analyzing its performance in comparison with the techniques studied previously.

\subsection{Analysis of the Membership Functions}


It is worthwhile considering that different datasets will have different frequency distributions for each criterion. Few terms in a collection tend to be in many documents, while many terms are used seldom. The effect of normalizing frequencies with respect to the most frequent term is that low values are compressed, and hence under-represented. This compression effect would exacerbate if the total maximum of the collection was used for the normalization process.

The fuzzy sets for FCC and EFCC were symmetrically defined, except for emphasis. Thus, some of the fuzzy sets defined for FCC and EFCC would match the initial state of most of the tuning processes of fuzzy rule-based systems.


In fact, what we call \emph{high} or \emph{low} are not absolute, but relative values. Therefore, a term is considered important because its normalized frequency is higher than most of the rest, and a certain value being \emph{high}, \emph{medium} or \emph{low} depends on the frequency distribution of the dataset. In an ideal case, all term frequencies would be uniformly distributed between $0$ and $1$ (see Figure \ref{sf:frequency-sets}), configuring the basic parameters of the fuzzy set using the original heuristic information. However, the fact that texts tend to follow Zipf's law, suggests that the uniform distribution is not always the case and more sophisticated approaches are needed. Hence, we believe that each particular dataset should have its own features and tuning of membership functions.



\subsection{Tuning of the Membership Functions}

Given the limitations of FCC, EFCC and AddFCC to deal with varying term distributions across different datasets, we now delve into alternative considerations that further exploit these characteristics, which ultimately leads to the definition of AFCC. In order to automatically adjust the basic parameters of the membership functions, we assume the two base cases that both the words in the documents as well as the emphasized terms will approximate a Zipfian distribution, as defined by Zipf's law \cite{zipf1949human}. For the first base case based on the \texttt{frequency} criterion, we consider we have a distribution tending to a power law when the majority of terms, i.e., more than a half of them ($55\%$) have normalized frequencies below $0.2$. Depending on whether this condition is fulfilled or not, we set the membership functions with one of the following two alternatives: 

\begin{enumerate}
\item When the precondition is fulfilled, we assume a distribution tending to a power law. As we need $5$ intervals to build three sets (\emph{low}, \emph{medium} and \emph{high} and two intersection areas between them, see Figure \ref{sf:frequency-sets}), our worst case would be to have only one possible value for each interval, that is, a maximum frequency of $5$. Thus, to guarantee at least one possible value for the low set in that case, we chose the first interval from $0$ to $1/5$. The rest of the intervals are selected using equidistant percentiles for term frequencies from $1/5$ to $1$, because this is suitable for the normalized frequencies that we found in our test data; 

\item When our precondition is not fulfilled, then we assume that the distribution tends to be closer to uniform, so that we can establish the fuzzy sets with the original heuristic, that is, all of them will have the same size. We use the corresponding percentiles to fit the distribution slightly better than using exact values ($0.2$, $0.4$, $0.6$, $0.8$, see Figure \ref{sf:frequency-sets}). Notice that in case of a uniform distribution, the adjustment for the first case---distribution tending to Zipf's law---would lead to these exact values too, because as the distribution moves towards a uniform distribution, the percentile $0.2$ will approximate to $1/5$ and the rest of the parameters belong to equidistant percentiles relative to this initial value in both cases. In those cases, the fuzzy sets would be symmetrical, that is, not only the case of the original sets in FCC and EFCC, but also the initial case used by most of the tuning methods of fuzzy rule-based systems.
\end{enumerate}

With regard to the \texttt{emphasis} criterion, we follow the same precondition as with \texttt{frequency} to determine whether or not the distribution tends to a power law, but modifying the fitting rules due to the different meaning of emphasis.
Again, we have two alternatives for \texttt{emphasis}: 

\begin{enumerate}
\item When the distribution tends to a power law, we set the first interval as in the \texttt{frequency} case, and the rest with decreasing percentiles, each being a half of the previous. The reason is that in the original heuristic-based fuzzy sets, the medium set is the biggest one, and we want to preserve the original heuristic knowledge, but always taking into account the relative difference between the number of elements in each set instead of absolute exact values;

\item If the collection does not fulfill our precondition, we assume that the distribution tends to be more uniform, so that we can establish the basic parameters of the membership functions by using the original heuristic rules but, as in the case of \texttt{frequency}, we use the percentiles instead of the exact values to fit slightly better the distribution (in this case the values were $0.05$, $0.15$, $0.55$, $0.75$, see Figure \ref{sf:emphasis-sets}).
\end{enumerate}

In the case of titles, we use the lowest value of the distribution to set the first interval, dividing the rest of the space in equidistant percentiles. Finally, it must be noted that we do not adjust the auxiliary system because the positions of words in a page do not depend on anything else than the number of words in the document.


We refer to this new approach we came up with after the analysis and tuning of the membership functions as Abstract Fuzzy Combination of Criteria (AFCC), which we test and evaluate next.

\subsection{Empirical Analysis of AFCC}

As AFCC represents a modification over the fuzzy logic based approaches, we use FCC and EFCC as baselines, as well as TF-IDF.
We apply the MFT reduction in all cases to compare the weighing functions in the same conditions.

\begin{figure}
 \centering
 \caption{Graphical representation of data in Table \ref{table:afcc_full}.}
 \includegraphics[width=\textwidth]{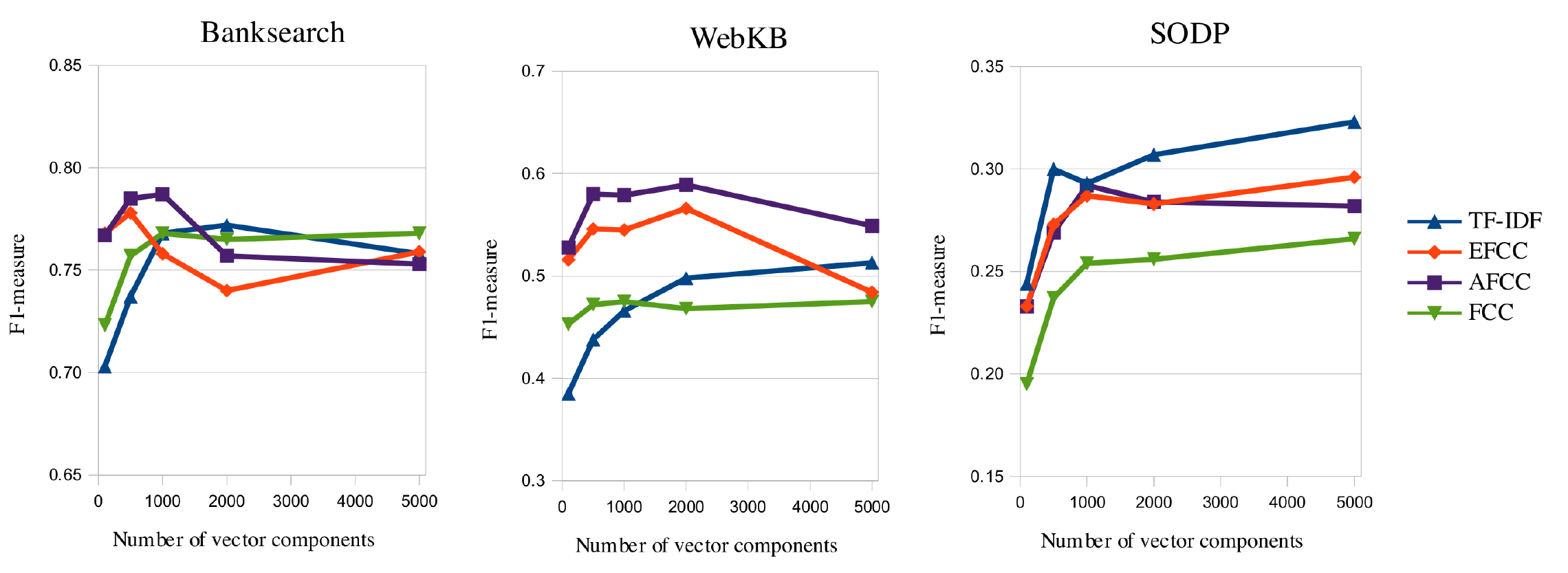}
 \label{fig:afcc_full}

 \footnotesize
 \begin{tabular}{| l | r | r | r | r | r | r |}
  \hline
  \textbf{Rep.} & \textbf{100} & \textbf{500} & \textbf{1000} & \textbf{2000} & \textbf{5000} & \textbf{Avg.} \\ 
  \hline 
  \multicolumn{7}{| c |}{\textbf{Banksearch}} \\
  \hline
  TF-IDF & 0.703 & 0.737 & 0.768 & \textbf{0.772} & 0.758 & 0.748 \\ 
  FCC & 0.723 & 0.757 & 0.768 & 0.765 & \textbf{0.768} & 0.756 \\
  EFCC & \textbf{0.768} & 0.778 & 0.758 & 0.740 & 0.759 & 0.760 \\
  AFCC & 0.767 & \textbf{0.785} & \textbf{0.787} & 0.757 & 0.753 & \textbf{0.770} \\
  \hline
  \multicolumn{7}{| c |}{\textbf{WebKB}} \\ 
  \hline 
  TF-IDF & 0.385 & 0.438 & 0.466 & 0.498 & 0.513 & 0.460 \\ 
  FCC & 0.453 & 0.472 & 0.475 & 0.468 & 0.475 & 0.469 \\ 
  EFCC & 0.516 & 0.546 & 0.545 & 0.566 & 0.484 & 0.532 \\ 
  AFCC & \textbf{0.528} & \textbf{0.580} & \textbf{0.579} & \textbf{0.589} & \textbf{0.549} & \textbf{0.565} \\ 
  \hline
  \multicolumn{7}{| c |}{\textbf{SODP}} \\ 
  \hline 
  TF-IDF & \textbf{0.244} & \textbf{0.300} & \textbf{0.293} & \textbf{0.307} & \textbf{0.323} & \textbf{0.293} \\
  FCC & 0.195 & 0.237 & 0.254 & 0.256 & 0.266 & 0.242 \\
  EFCC & 0.233 & 0.273 & 0.287 & 0.283 & 0.296 & 0.275 \\ 
  AFCC & 0.233 & 0.269 & 0.292 & 0.284 & 0.282 & 0.272 \\ 
  \hline
 \end{tabular}
 \captionof{table}{F1 results for membership functions experiments (all with MFT reduction).}
 \label{table:afcc_full}
\end{figure}

Table \ref{table:afcc_full} and Figure \ref{fig:afcc_full} show F1 scores for these representations. On the one hand, looking at the results, among the fuzzy logic based representations, AFCC outperforms the rest in WebKB in all cases, while in Banksearch got better results than the others with $2$ out of $5$ vector sizes, having also a higher average F1 score. This varying performance across collections could be due to the fact that frequency distributions in Banksearch rather approximate a power law. In those cases, the least frequent terms are assigned to the low fuzzy set, with few terms remaining for the medium and high sets. This explains the small difference between the EFCC and FCC fixed sets. The same occurs with SODP, where EFCC and AFCC get similar results, although FCC performs worse, probably due to its underestimation of \texttt{frequency}. However, with a rather uniform term frequency distribution, as in WebKB, adjusting the fuzzy sets has a much bigger effect in results, where more terms are assigned to the medium and high fuzzy sets, and small variations of the basic parameters of the membership functions will have a much bigger effect. It is indeed important to adapt to this kind of distributions, as the terms are differently used and structured.

On the other hand, TF-IDF obtained surprisingly good results in SODP compared to the results of the same function with Banksearch and WebKB datasets. In general, results with all the representations tend to be worse in SODP, due to the special difficulties of this collection. We believe that the use of IDF could help improve the results of TF-IDF because it would alleviate the effect of the bigger categories, whose terms would be penalized giving more representativeness to those belonging to smaller categories. This fact would reduce slightly the bias introduced by the bigger categories, allowing to cluster the smaller ones slightly better. This improvement in the clustering of smaller categories could lead to an improvement in the overall clustering results of TF-IDF.

\begin{sidewaysfigure*}[p]
 \begin{center}
   \includegraphics[width=1.0\textwidth]{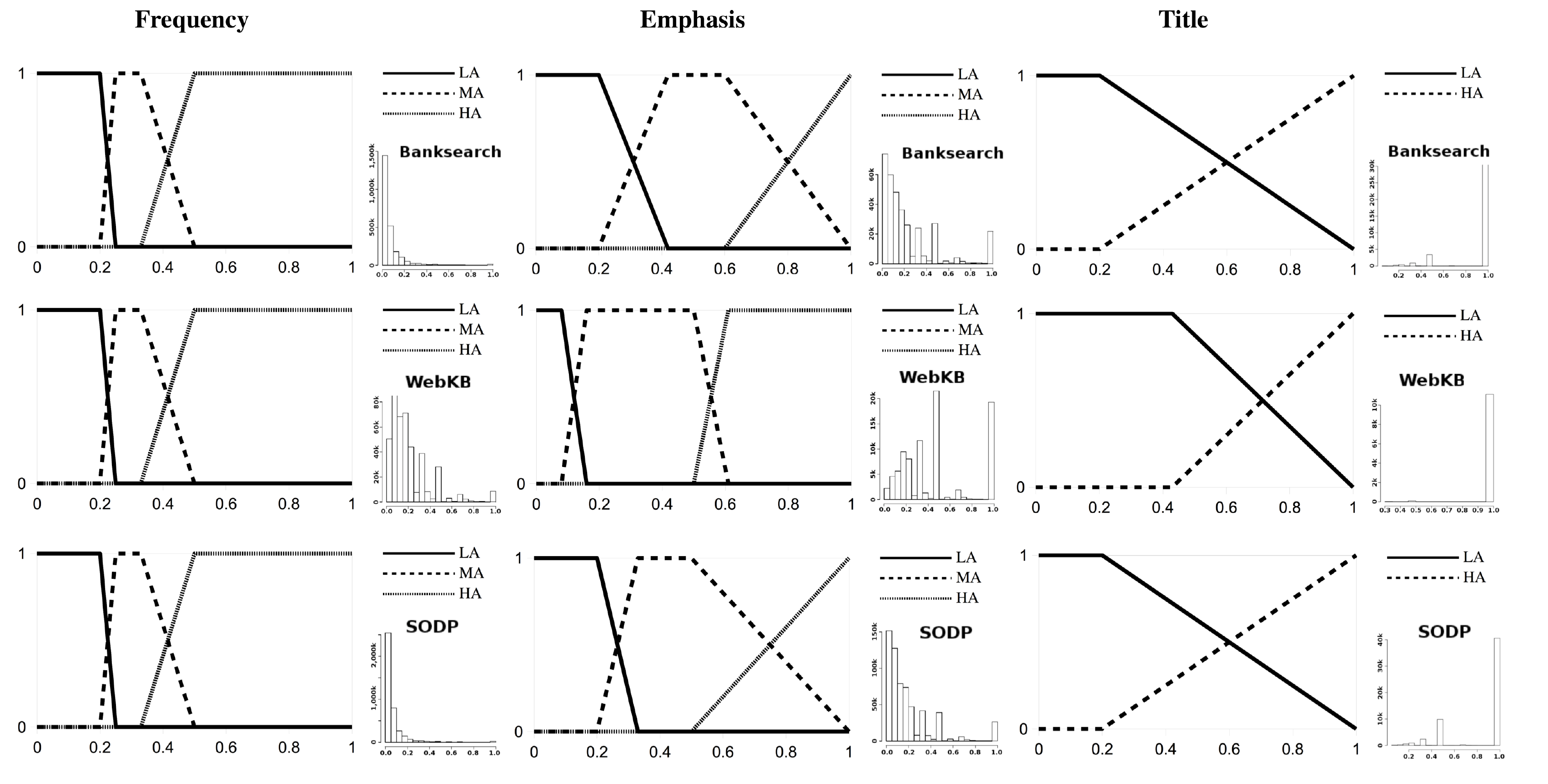}
   \caption{Membership functions and distributions of input values for each criterion and dataset. Columns represent criteria, while rows represent datasets. Each pair of dataset and criterion includes two charts. The left-hand side, bigger chart, shows the final fuzzy sets for the membership functions after the automatic adjustment. The right-hand side, smaller chart, shows the distribution of term frequencies normalized by document in the dataset for each criterion, where X axis refers to the input value for the criterion, and Y axis refers to the number of terms that belong to that bin.}
   \label{fig:membership-functions}
 \end{center}
\end{sidewaysfigure*}

In general, adjusting the membership functions to the dataset seems to be useful not only to add more automatism to the document representation process, but also because this automation allows the system to adapt better to datasets with specific characteristics. The proposed method is able to achieve similar results to EFCC when dealing with exponential distributions. Moreover, when the shape of the distribution changes, the adjustment helps improve clustering results, as is the case of WebKB. Figure \ref{fig:membership-functions} shows a summary of the resulting membership functions and the distributions of input values for each dataset and criterion.

\subsection{Statistical significance}
\label{ss:afcc_significance}

We analyze in depth the difference between using membership function tuning and the original representation with fixed fuzzy sets. Besides, we also include FCC in the comparison. We are interested in seeing the global improvements of the new proposal, AFCC, with respect to the original baseline. Each dataset was divided in 100 different sub- datasets 50\% smaller than the original, where the size of each category is in proportion to the original ones. 
We performed 100 experiments per vector size corresponding to each sub-dataset, resulting in a total of 4,500 different clustering experiments. We calculated the statistical significance between F1 scores of each pair of representations (AFCC-FCC, EFCC-FCC, AFCC-EFCC) with a paired two-tailed t-test for each vector size.

\begin{figure}
 \centering
 \caption{Graphical representation of data in Table \ref{table:afcc_ttest}.}
 \includegraphics[width=\textwidth]{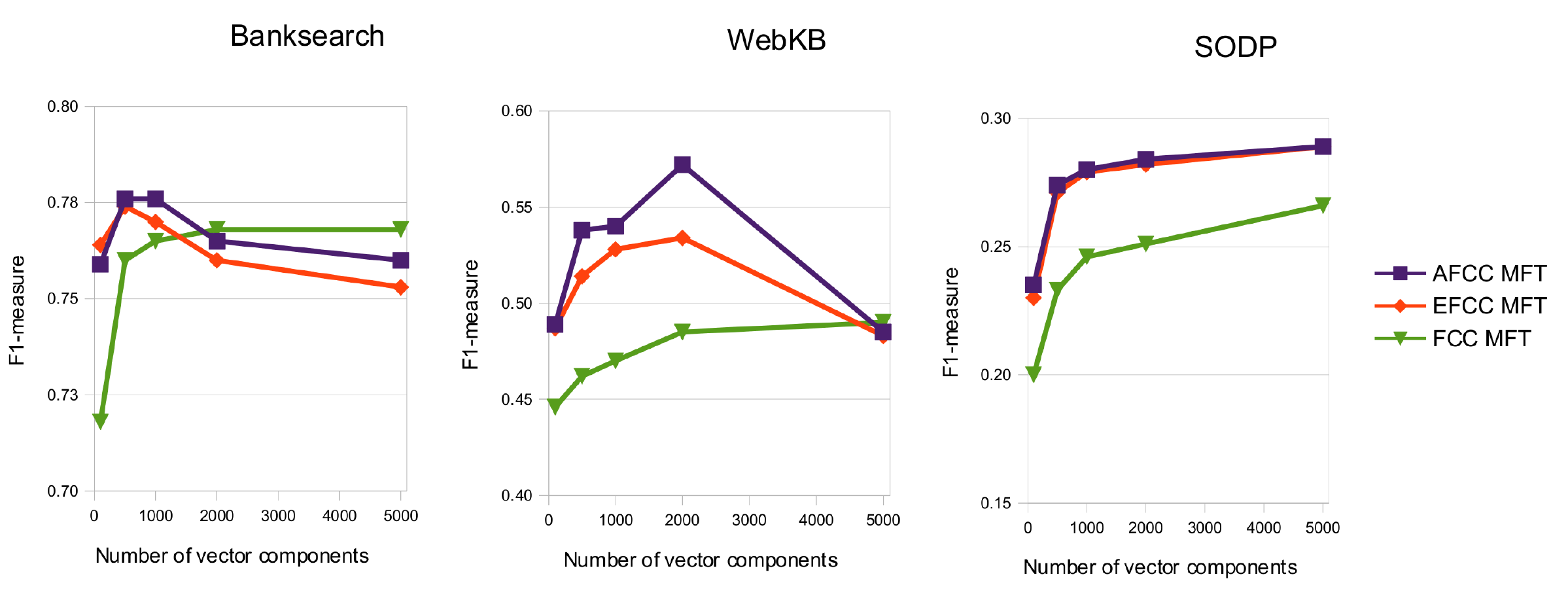}
 \label{fig:afcc_ttest}

 \footnotesize
 \begin{tabular}{lllllll}
  \hline
  &  & \textbf{100} & \textbf{500} & \textbf{1000} & \textbf{2000} & \textbf{5000} \\ 
  \hline
  \multicolumn{6}{c}{\textbf{Banksearch}} \\
  \hline
  \multirow{3}{15mm}{F1-measure} 
  & AFCC MFT & 0.759 & \textbf{0.776} & \textbf{0.776} & 0.765 & 0.760 \\ 
  & EFCC MFT & \textbf{0.764} & 0.774 & 0.770 & 0.760 & 0.753 \\
  & FCC MFT & 0.718 & 0.760 & 0.765 & \textbf{0.768} & \textbf{0.768} \\ 
  \hline
  \multirow{3}{15mm}{Difference} 
  & AFCC-FCC & 0.041$^{**}$ & 0.016$^{**}$ & 0.011$^{**}$ & -0.003 & -0.007$^{**}$\\ 
  & EFCC-FCC & 0.047$^{**}$ & 0.014$^{**}$ & 0.006$^{*}$ & -0.008$^{**}$  & -0.015$^{**}$\\ 
  & AFCC-EFCC & -0.005$^{**}$ & 0.003 & 0.005$^{*}$ & 0.005$^{**}$  & 0.008$^{*}$\\
  \hline
  \multicolumn{6}{c}{\textbf{WebKB}} \\ 
  \hline 
  \multirow{3}{15mm}{F1-measure} 
  & AFCC MFT & \textbf{0.489} & \textbf{0.538} & \textbf{0.540} & \textbf{0.572} & 0.485 \\
  & EFCC MFT & 0.487 & 0.514 & 0.528 & 0.534 & 0.483 \\
  & FCC MFT & 0.446 & 0.462 & 0.470 & 0.485 & \textbf{0.490} \\
  \hline
  \multirow{3}{15mm}{Difference} 
  & AFCC-FCC & 0.043$^{**}$ & 0.076$^{**}$ & 0.070$^{**}$ & 0.087$^{**}$  & -0.004\\
  & EFCC-FCC & 0.041$^{**}$ & 0.051$^{**}$ & 0.059$^{**}$ & 0.049$^{**}$  & -0.007\\ 
  & AFCC-EFCC & 0.002 & 0.025$^{**}$ & 0.011$^{**}$ & 0.038$^{**}$  & 0.002$^{*}$\\
  \hline
  \multicolumn{6}{c}{\textbf{SODP}} \\ 
  \hline 
  \multirow{3}{15mm}{F1-measure} 
  & AFCC MFT & \textbf{0.235} & \textbf{0.274} & \textbf{0.280} & \textbf{0.284} & \textbf{0.289} \\
  & EFCC MFT & 0.230 & 0.271 & 0.279 & 0.282 & \textbf{0.289} \\
  & FCC MFT & 0.200 & 0.233 & 0.246 & 0.251 & 0.266 \\
  \hline
  \multirow{3}{15mm}{Difference} 
  & AFCC-FCC & 0.035$^{**}$ & 0.040$^{**}$ & 0.033$^{**}$ & 0.033$^{**}$  & 0.023$^{**}$\\
  & EFCC-FCC & 0.030$^{**}$ & 0.037$^{**}$ & 0.033$^{**}$ & 0.031$^{**}$  & 0.023$^{**}$\\ 
  & AFCC-EFCC & 0.005$^{**}$ & 0.003$^{*}$ & 0.000 & 0.001  & 0.000\\
  \hline
 \end{tabular}
 \captionof{table}{Results for AFCC/EFCC/FCC t-test experiments. $^{*}$ indicates a statistically significant difference at $p < 0.01$, and $^{**}$ indicates a statistically significant difference at $p < 0.001$.}
 \label{table:afcc_ttest}
\end{figure}

In Table \ref{table:afcc_ttest} and Figure \ref{fig:afcc_ttest}, for each vector size and representation we show the average F1 scores of the 100 clustering experiments (one per sub-dataset), and in Table \ref{table:afcc_ttest} we also show the difference between the corresponding averages, and the \emph{p}-value resulting from applying the statistical t-test between each pair of representations. 

In most of the cases AFCC outperforms EFCC, and consequently also FCC. Therefore, the difference between term frequency distributions of the datasets, in combination with all of these results allow us to conclude that membership function tuning helps determine each criterion in a better way, ultimately improving clustering results.


Adjusting the membership functions to a dataset leads to results as good as or better than FCC in $91.6\%$ of the cases, and as good as or better than EFCC in $86.5\%$ of the cases. EFCC and AFCC outperform FCC in most of the cases, and between them, AFCC allows to improve the results of EFCC in $28.5\%$ of the cases.

\section{Discussion}
\label{sec:discussion}

We have studied the application of fuzzy logic for the representation of web documents in a way that imitates humans skimming through the documents. The use of fuzzy logic enables us to separate the knowledge declaration from the calculation procedure, which also enables us to specify the knowledge by means of a set of rules close to natural language applying non linear combinations of criteria. Building on a state-of-the-art unsupervised document representation, Fuzzy Combinations of Criteria (FCC), we have introduced, evaluated, and analyzed three alternatives that make the most of the HTML structure and content of web documents, namely EFCC, AddFCC, and AFCC. We evaluate and compare the representation approaches in a web page clustering task, using three datasets with very different characteristics.

We first defined a set of rules fixed on the basis of expert knowledge. Although there are other options to automatically generate these rules (the rule base could be adjusted by using machine learning techniques that adapt sets of rules to sets of sample data, or by using bio-inspired approaches), both approximations could cause a loss of generality in the learned/generated model in the attempt to fit the system to specific sample data. This could lead to illogical rules. On the other hand, in this automated scenario, we would need to deal with the coherence of the rules, which would require to establish a methodology to measure this coherence among rules. Last but not least, our system evaluates each term within each document using a fuzzy approach, which implies a high computational cost. Therefore, the use of machine learning or bio-inspired algorithms would add a considerable cost to the system. Of course, the manual definition of the rules employed in our work could lead to mistakes in the knowledge definition. However, in the same way, the application of machine learning or bio-inspired techniques would always require an initial knowledge to start the process.

Considering these aspects, we analyzed three challenges concerning web page representation for clustering: (1) the selection of feature sources to extract essential information from; (2) the term weighing functions to estimate the weight of each feature; and (3) the dimensionality reduction techniques to select the most representative features and to reduce the computational cost of the clustering. 

For feature selection, we explored the application of new, mostly unstudied criteria to improve the representation with information from the whole collection as well as from anchor texts. 
For term weighing we explored the fuzzy combination of criteria performed by FCC \cite{Fresno06} aiming to get the most of the fuzzy system and the heuristics in which it is based. We use TF-IDF as the baseline, since it is a standard weighting method employed to represent documents. We presented an improved representation called EFCC, which outperformed the baselines, and another alternative called AddFCC, which did not work as well as expected. Both alternatives attempt to exploit the fuzzy system in a different manner to FCC, taking advantage of its additive properties. 
For dimensionality reduction, we introduced MFT, a lightweight dimensionality reduction technique, based on the term weighing function, which is able to improve the results of more complex techniques such as LSI when used together with EFCC in our test collections.

We also studied whether EFCC could be tuned to fit the specific characteristics of different collections. The aim of this adjustment is not only to improve clustering results in those collections, but also to adapt the representation to different datasets that could have different features. We found the case of the WebKB dataset, which has very different characteristics, particularly when looking at terms that are emphasized within the document contents. This led us to further study the tuning of the fuzzy system in an unsupervised way, for which we proposed AFCC. AFCC adjusts the basic parameters of the membership function on the basis of the term distributions of the collections. We showed that AFCC leveled or even improved the good results of EFCC and FCC in all kinds of datasets, outperforming the results of other approaches. Our results show that AFCC is a competitive approach that outperforms the rest of the techniques, with good performance across datasets of very different characteristics.

Future work includes the study of the effect of non-linear scaling factors as a complementary tool to our proposal to adjust the representation to specific datasets, and to study new ways of considering the position criterion. A complementary analysis would include the exploitation of the position of words in documents through visual rendering of web pages. Finally, it would be interesting to study and assess the inclusion of additional criteria in the combination.

\ifCLASSOPTIONcompsoc
  \section*{Acknowledgments}
\else
  \section*{Acknowledgment}
\fi

The authors would like to thank the anonymous reviewers for their valuable comments and suggestions to improve the quality of the paper.
This work has been part-funded by the Spanish Ministry of Science and Innovation (MED-RECORD Project, TIN2013-46616-C2-2-R) and the PHEME FP7 project (grant No. 611233).

\bibliographystyle{IEEEtran}
\bibliography{alp}

\end{document}